\documentclass[aps, prd, preprintnumbers, letterpaper, nofootinbib, superscriptaddress, balancelastpage, 12pt, reprint, onecolumn]{revtex4-2}
\usepackage{babel}
\usepackage[utf8]{inputenc}
\usepackage[T1]{fontenc}
\usepackage{graphicx}
\usepackage{xcolor}
\usepackage{amsmath, amsfonts, amssymb}
\usepackage{siunitx}
\usepackage[caption=false]{subfig}
\def\linkcolor{blue!90!black}
\usepackage[
  colorlinks=true,
  urlcolor=\linkcolor,
  anchorcolor=\linkcolor,
  citecolor=\linkcolor,
  filecolor=\linkcolor,
  linkcolor=\linkcolor,
  menucolor=\linkcolor,
  linktocpage=true
]{hyperref}

\usepackage[nameinlink,capitalize]{cleveref}

\graphicspath{{./figures/}}
\newcommand{\dd}{\mathrm{d}}

\begin{document}
%%%%%%%%%%%%%%%%%%%%%%%%%%%%%%%%
%%%%%%%%%%%%%%%%%%%%%%%%%%%%%%%%

\begin{minipage}{18cm}
\vspace{-1cm}
    \begin{flushright}
DESY-25-144\\
IFT-UAM/CSIC-25-118
\end{flushright}
\end{minipage}

\title{
Gravitational waves from the sound shell model:\texorpdfstring{\\}{} direct and inverse phase transitions in the early Universe
}
\author{Giulio Barni}
 \email{giulio.barni@ift.csic.es}
\affiliation{Instituto de F\'isica Te\'orica IFT-UAM/CSIC, Cantoblanco, E-28049, Madrid, Spain}
\author{Simone Blasi}
\email{simone.blasi@desy.de}
\affiliation{Deutsches Elektronen-Synchrotron DESY, Notkestr.~85, 22607 Hamburg, Germany}
\author{Eric Madge}
\email{eric.madgepimentel@uam.es}
\affiliation{Instituto de F\'isica Te\'orica IFT-UAM/CSIC, Cantoblanco, E-28049, Madrid, Spain}
\affiliation{Departamento de F\'isica Te\'orica, Universidad Aut\'onoma de Madrid, Cantoblanco, E-28049 Madrid, Spain}
\author{Miguel Vanvlasselaer}
\email{miguel.vanvlasselaer@ub.edu}
\affiliation{Theoretische Natuurkunde and IIHE/ELEM, Vrije Universiteit Brussel,
\& The International Solvay Institutes, Pleinlaan 2, B-1050 Brussels, Belgium}
\affiliation{Departament de Física Quàntica i Astrofísica and Institut de Ciències del Cosmos (ICC), 
Universitat de Barcelona, Martí i Franquès 1, ES-08028, Barcelona, Spain.}

\begin{abstract}
Cosmological first order phase transitions are a frequent phenomenon in particle physics beyond the Standard Model, and the corresponding gravitational wave signal offers a key probe of new physics in the early Universe. Depending on the underlying microphysics, the transition can exhibit either direct or inverse hydrodynamics, leading to a different phenomenology. Most studies to date have focused on direct transitions, where the cosmic fluid is pushed or dragged by the expanding vacuum bubbles. In contrast, inverse phase transitions are characterized by fluid profiles where the plasma is sucked in by the expanding bubbles. 
Using the sound shell model, we derive and compare the gravitational wave spectra from sound waves for direct and inverse phase transitions, providing new insights into the potential observable features and the possibility of discriminating among the various fluid solutions in gravitational wave experiments.
\end{abstract}

\maketitle

\section{Introduction}

We study cosmological phase transitions~(PTs) in the early Universe, with a focus on first-order phase transitions (FOPTs) that proceed via the nucleation and expansion of bubbles of the true vacuum within a sea of false vacuum; these transitions can trigger baryogenesis~\cite{Kuzmin:1985mm, Shaposhnikov:1986jp, Nelson:1991ab, Carena:1996wj, Cline:2017jvp, Long:2017rdo, Bruggisser:2018mrt, Bruggisser:2018mus, Bruggisser:2022rdm, Morrissey:2012db, Azatov:2021irb, Huang:2022vkf, Baldes:2021vyz, Chun:2023ezg, Cataldi:2024pgt, Cataldi:2025nac}, lead to the production of dark matter~\cite{Falkowski:2012fb, Baldes:2020kam, Hong:2020est, Azatov:2021ifm, Baldes:2021aph, Asadi:2021pwo, Lu:2022paj, Baldes:2022oev, Azatov:2022tii, Baldes:2023fsp, Kierkla:2022odc, Giudice:2024tcp}, generate primordial black holes~\cite{Kodama:1982sf, Kawana:2021tde, Jung:2021mku, Gouttenoire:2023naa, Lewicki:2023ioy}, and source stochastic gravitational wave~(GW) backgrounds~\cite{Witten:1984rs, Hogan:1986qda, Kosowsky:1992vn, Kosowsky:1992rz, Kamionkowski:1993fg}. The discovery prospects with space- and ground-based next-generation GW interferometers such as LISA, ET, CE, BBO, and DECIGO open a unique window onto otherwise inaccessible sectors of particle physics, as many well-motivated Beyond the Standard Model (BSM) frameworks naturally realize cosmological PTs, including composite Higgs models~\cite{Pasechnik:2023hwv, Azatov:2020nbe, Frandsen:2023vhu, Reichert:2022naa, Fujikura:2023fbi}, extended Higgs sectors~\cite{Delaunay:2007wb, Kurup:2017dzf, VonHarling:2017yew, Azatov:2019png, Ghosh:2020ipy, Aoki:2021oez, Badziak:2022ltm, Blasi:2022woz, Agrawal:2023cgp, Banerjee:2024qiu}, axion models~\cite{DelleRose:2019pgi, VonHarling:2019rgb}, dark sectors~\cite{Schwaller:2015tja, Breitbach:2018ddu, Fairbairn:2019xog, Halverson:2020xpg, Morgante:2022zvc, Ghosh:2022fzp, Hooper:2025fda}, and $B\!-\!L$-breaking sectors~\cite{Jinno:2016knw, Addazi:2023ftv}.

Against this backdrop, interest in \emph{inverse} phase transitions has grown, particularly in cosmological settings that feature transient heating phases --- for example, reheating after inflation or from energy injection\,\cite{Buen-Abad:2023hex,Barni:2025ced,Dent:2024bhi,Dent:2025bwo,Sui:2025szm}, regions influenced by the rarefaction tails of bubbles from a preceding (direct) phase transition \cite{Caprini:2011uz}, and localized heating near primordial black holes (PBHs) \cite{Ai:2024cka} and in the core of dense neutron stars~\cite{Casalderrey-Solana:2022rrn, Bea:2024bxu, Bea:2024bls} --- in which the plasma’s thermodynamic trajectory can favor an inverse transition and thereby alter the standard, monotonic-cooling picture. The hydrodynamics of inverse phase transitions has been investigated in~\cite{Barni:2024lkj, Bea:2024bxu}, highlighting how heating backgrounds and nontrivial flow profiles modify the interplay between bubble expansion, shock/rarefaction fronts, and energy-momentum transport. Perhaps even more surprisingly, symmetry-breaking phase transitions while standard cosmological cooling have recently been shown to arise within supersymmetric models~\cite{Barni:2025mud} and BSM realisations of the EWPT \cite{Ai:2025vfi}, adding theoretical motivation to revisit and systematize inverse-transition dynamics.

Despite these advances, a comprehensive analysis of the gravitational-wave signal sourced by inverse phase transitions --- together with a more complete study of the associated bubble-wall velocity, including its microphysical inputs and hydrodynamic constraints --- has so far been lacking. In our view, the most direct and reliable way to close this gap would be a fully coupled numerical simulation that evolves both the scalar field and the relativistic fluid, since the GW source depends on the stress-energy of \emph{both} sectors and on their mutual backreaction. Recent progress toward such end-to-end simulations for strong first-order transitions shows the power of this approach~\cite{Correia:2025qif}. In contrast, simulations that evolve only the fluid, such as the “Higgsless” setup~\cite{Jinno:2022mie}, allow to efficiently explore the different hydrodynamic regimes while accounting for a large statistics in the number of bubbles but, by construction, cannot capture the full set of effects that arise from the interplay between field gradients, wall microphysics, and hydrodynamics. 

Until field-fluid simulations tailored to inverse transitions are available, a pragmatic first step is i) to devise a reliable framework for computing the bubble wall velocity during inverse phase transitions using the \emph{local thermal equilibrium} (LTE) approximation, ii) to quantify the sound-wave contribution during and after bubble collisions within the sound shell model (SSM), which is commonly used as a reference framework for comparison with numerical simulations.

The LTE approximation~\cite{Balaji:2020yrx, Ai:2021kak, Ai:2023see,Ai:2024shx} is a simple and flexible assumption customarily used to obtain an estimate of the wall velocity~\cite{Ai:2024btx}.
Within this approximation, the plasma on both sides and inside the bubble wall is assumed to remain in equilibrium, with entropy conservation linking the temperatures as $T_{-}/T_{+} = \gamma_{+}/\gamma_{-}$, where subscripts `$\pm$' denote quantities on the outer (`$+$') and inner (`$-$') side of the wall, and $\gamma_\pm$ are the boost factor of the fluid velocity on either side of the wall (in the wall frame). Under this assumption, the hydrodynamic equations reduce to a one-parameter family of solutions that can be fully determined. The matching conditions across the wall relate the fluid velocities and temperatures without having to solve the Boltzmann equation explicitly. This approximation then provides a controlled and efficient way to estimate the wall dynamics and the resulting hydrodynamic profiles driving the GW source.

The SSM provides a physically transparent description of how overlapping spherical sound shells generate a stochastic background of GWs within the approximation of linear and Gaussian dynamics. Originally proposed in Refs.~\cite{Hindmarsh:2016lnk,Hindmarsh:2019phv}, it has been extended to include e.g.\ the expansion history of the Universe and different equations of state~\cite{Guo:2020grp,RoperPol:2023dzg,Giombi:2024kju,Giombi:2025tkv} and used in observational studies~\cite{Gowling:2021gcy,Boileau:2022ter,Gowling:2022pzb,Guo:2024gmu,Tian:2025zlo}. In this work we adopt the SSM, closely following the implementation in Ref.~\cite{RoperPol:2023dzg}, as our baseline framework to capture the dominant acoustic contribution for both direct and inverse transitions.
It is important to emphasize that the qualitative features unique to inverse transitions, most notably the inflow fluid dynamics across the bubble wall, call for dedicated numerical simulations as a crucial next step toward accurate gravitational-wave predictions.

The remainder of this paper is organised as follows: in \cref{sec:reminder}, we remind the salient properties of direct and inverse phase transitions; in \cref{sec:velocity}, we study the velocity of the bubble wall for the direct and inverse phase transitions using the LTE approximation; in \cref{sec:sound_shell} we use the SSM to estimate the spectrum of GWs emitted by a direct and inverse phase transitions and discuss the salient differences between the GW signal induced by the direct and the inverse PTs. In \cref{sec:results}, we present the results of the analysis of the GW signal with the SSM model for the direct and inverse phase transitions.  We conclude in \cref{sec:conclusion}. 

\medskip
Our code used to calculate the fluid profiles and gravitational wave signals for direct and inverse phase transitions is publicly available at \url{https://github.com/eric-madge/inverse_pt}.

\section{Reminder on inverse phase transitions}
\label{sec:reminder}

\noindent Before turning to hydrodynamics, we clarify the distinction between \emph{direct} and \emph{inverse} first-order phase transitions in terms of the \emph{sign of the generalised pseudotrace}, $\alpha_\vartheta$ (cf.\ \cref{eq:alpha_general}). In Ref.~\cite{Barni:2025mud} it was observed that the quantity that most clearly distinguishes the different hydrodynamic solutions is a suitable generalisation of the trace anomaly of the stress-energy tensor, namely generalized pseudotrace $\alpha_\vartheta$. To make contact with more familiar thermodynamic quantities, it was shown that this object coincides with the standard thermodynamic latent heat $L\equiv\Delta w(T_c)$ when evaluated at the critical temperature. It is therefore natural to interpret $\alpha_\vartheta$ as the effective latent heat of the phase transition computed at the nucleation temperature rather than at the critical one. Then:
\begin{itemize}
  \item \textbf{Direct PT (\(\alpha_\vartheta>0\)).} The transition \emph{releases} energy into the environment. The injected energy drives an outward pressure gradient in the plasma, sourcing fluid outflows from the advancing bubble wall, going ``away'' as seen from the center of the bubble.
  \item \textbf{Inverse PT (\(\alpha_\vartheta<0\)).} The transition \emph{absorbs} energy from the environment. Energy is drawn from the surrounding plasma into the bubble interface, which tends to establish inward pressure gradients and correspondingly favors fluid motion toward the center of the bubble.
\end{itemize}
In this description, “direct” versus “inverse” refers solely to the sign of \(\alpha_\vartheta\): positive \(\alpha_\vartheta\) implies energy release; negative \(\alpha_\vartheta\) implies energy uptake. Because inverse transitions require extracting heat from the medium, they are intrinsically thermal processes and \emph{cannot occur at} \(T=0\).

It is worth to briefly comment on the conceptual difference between \emph{inverse} and \emph{symmetry-restoring} first-order phase transitions. In a symmetry-restoring transition, the system evolves from a broken-symmetry phase to a symmetric one; from the symmetry point of view, it does not matter whether this happens upon cooling or heating of the plasma, as the notion of ``symmetry restoration'' only tracks the order parameter. By contrast, an \emph{inverse} phase transition is defined in hydrodynamical terms and is, in principle, independent of whether a symmetry is being restored or broken, and of whether the transition proceeds during cooling or heating. What matters is instead the sign of the generalised pseudotrace introduced above, which fixes the direction of the fluid flow: in the inverse case, the plasma is drawn \emph{into} the bubble rather than being pushed outwards. Examples of symmetry-breaking inverse phase transitions occurring upon cooling have been presented in Ref.~\cite{Barni:2025mud}, while examples of inverse symmetry-restoring phase transitions taking place during heating have been discussed in Ref.~\cite{Buen-Abad:2023hex,Barni:2025ced}. Notice that the phase transition presented in Ref.~\cite{Ai:2025vfi} was strictly speaking a symmetry-restoring direct phase transition.

This reversed fluid flow has profound implications for the dynamics of the transition. The hydrodynamic solutions that describe the plasma motion, such as detonations, deflagrations, and hybrids, are qualitatively modified, as the pressure and velocity profiles now correspond to an inward rather than outward propagation. 
From the point of view of the energy budget, ordinary (``direct'') first-order phase transitions are basically fueled by the $T=0$ vacuum energy difference between the two phases. For inverse transitions, instead, the situation is qualitatively different. At the nucleation temperature, the $T=0$ vacuum energy difference has the opposite sign, and needs to be interpreted as a \emph{product} of the transition rather than part of the initial energy budget. In this sense, it is the thermal contribution from the plasma that drives the bubble expansion.

The hydrodynamics of bubble growth follows from total energy-momentum conservation for the combined scalar-fluid system, $T^{\mu\nu}=T_\phi^{\mu\nu}+T_f^{\mu\nu}$. The scalar and fluid stress-energy tensors read
\begin{subequations}
\begin{align}
T_{\phi}^{\mu\nu}&=(\partial^\mu\phi)\partial^\nu\phi-g^{\mu\nu}\!\left(\tfrac{1}{2}(\partial\phi)^2-V(\phi)\right),\\
T_f^{\mu\nu}&=(e_f+p_f)u^\mu u^\nu- g^{\mu\nu}p_f,
\end{align}
\end{subequations}
where $u^\mu$ is the fluid four-velocity, and $e_f$ and $p_f$ are the thermal (vanishing at $T\!=\!0$) energy density and pressure. It is convenient to combine the thermal part with the vacuum potential energy by defining
\[
e \equiv e_f+V(\phi),\qquad p \equiv p_f - V(\phi),
\]
so that the enthalpy remains unchanged, $w\equiv e_f+p_f=e+p$, and the fluid tensor can be written as
\begin{align}
    T_f^{\mu\nu}=(e+p)u^\mu u^\nu- g^{\mu\nu}[p+V(\phi)].
\end{align}

The matching conditions follow from the $\nu=0$ and $\nu=3$ components of $\nabla_\mu T^{\mu\nu}=0$. Working in the instantaneous comoving frame of the bubble wall at time $\tau$, and writing $u^{\mu}=\gamma(z)\,(1,0,0,-v(z))$, one obtains \cite{Ai:2024shx}
\begin{subequations}
\label{eq:matching-original}
    \begin{align}
         \partial_z (w \gamma^2 v) &=0 \, ,\\
        -(\partial_z\phi)\,  \ddot{\phi}(0)\big|_{\tau}+\partial_z \!\left[w \gamma^2 v^2+\tfrac{1}{2}(\partial_z\phi)^2+p\right] &=0 \,. 
    \end{align}
\end{subequations}

To close the system we specify an equation of state (EoS). In the \emph{general} Bag model, one allows for (possibly different) vacuum energies $\epsilon_\pm$ in the two phases,
\begin{align}
\label{eq:bag_eos}
    &e_+(T)=a_+  T^4+\epsilon_+,\qquad p_+(T)=\tfrac{1}{3}a_+ T^4-\epsilon_+,\nonumber\\
    &e_-(T)=a_-  T^4+\epsilon_- ,\qquad\;\  p_-(T)=\tfrac{1}{3}a_- T^4-\epsilon_-, 
\end{align}
 where $\epsilon_\pm$ is the vacuum (zero-temperature) potential energy densities of the two phases and where a subscript ``$\pm$'' denotes quantities just in front of ($+$) and just behind ($-$) the wall. We also defined $a$, which encodes the number of \emph{relativistic} degrees of freedom in each phase, 
 \begin{equation}
    a_{\pm} = \frac{\pi^2}{30} \sum_{\text{light } i} 
\left[ N_i^b + \frac{7}{8} N_i^f \right] \, ,
\end{equation}
where $N^{b(f)}_i$ are the number of light bosons (fermions) dofs.  Explicitly, $w_+=w_+(T_+)$ and $w_-=w_-(T_-)$ (similarly for $p_{\pm}$). We stress that the phases $\pm$ could be associated to the symmetric/broken phase ($``s/b"$) in a symmetry-breaking phase transition.

Assuming vanishing first time-derivatives of the hydrodynamic variables, we can integrate \cref{eq:matching-original} across the wall to obtain\cite{Ai:2024shx}
\begin{subequations}
\label{eq:junctionAB}
\begin{align}
      w_+\gamma_+^2v_+ &=w_-\gamma_-^2v_-,\label{eq:conditionA}\\
      0 = -\!\int_{-\delta}^\delta \! \dd z\, (\partial_z\phi)\, \ddot{\phi}(0)\big|_{\tau} &=[w_-\gamma_-^2v_-^2+p_-]-[w_+\gamma_+^2v_+^2+p_+] \equiv \Delta V- \Delta\!\left(-V_T+w \gamma^2 v^2\right),\label{eq:conditionB}
\end{align}
\end{subequations}
where we \emph{define} $
\Delta V \;\equiv\; \epsilon_+ - \epsilon_- \, .
$

\begin{figure}[t]
  \hspace*{\fill}
  \includegraphics{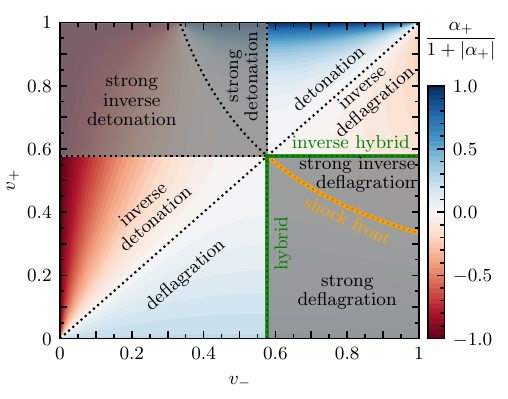}
  \hfill
  \includegraphics{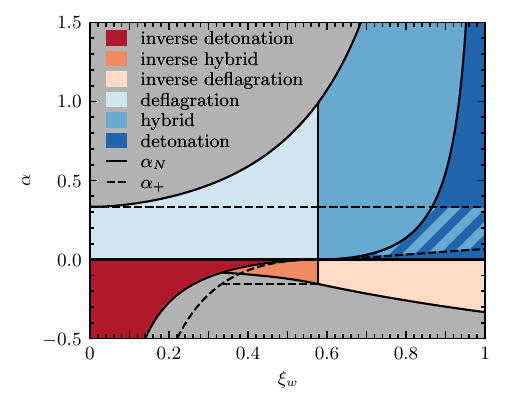}
  \hspace*{\fill}
  \caption{Hydrodynamic solution space for direct and inverse first-order phase transitions.
  \textbf{\boldmath Left (branch structure in the $(v_+,v_-)$ plane):} the two hydrodynamic branches from \cref{eq:vplus_vminus_relation} are shown in the wall frame with $v_-$ on the horizontal axis and $v_+$ on the vertical axis. 
  The color coding indicates the corresponding value of the transition strength parameter $\bar{\alpha}_+$ normalized to the range $[-1,1]$, $\bar{\alpha}_+=\alpha_+/(1+|\alpha_+|)$, showing inverse (negative $\bar{\alpha}_+$) and direct (positive $\bar{\alpha}_+$) transitions in red and blue, respectively.
  The locus annotated as ``shock front'' marks the region where a shock forms ahead of the wall in deflagrations, consistent with the standard classification into deflagration, hybrid, and detonation (and their inverse counterparts). 
  \textbf{\boldmath Right (classification in the $(\alpha,\xi_w)$ plane):} the horizontal axis is the wall speed $\xi_w$ in the plasma rest frame; the vertical axis is the strength $\alpha$. Solid curves refer to contours/boundaries for $\alpha_N$ (evaluated at $T_N$), while dashed curves refer to $\alpha_+$ (evaluated just ahead of the wall). For a given point $(\alpha_N,\xi_w)$ the solution is unique, whereas for $(\alpha_+,\xi_w)$ there exists a region (hatched) in which two distinct hydrodynamic solutions (detonation or hybrid) share the same pair $(\alpha_+,\xi_w)$. Colors follow the legend in the panel; the shaded gray area denotes kinematically forbidden configurations.}
  \label{fig:hydro_solution_space}
\end{figure}

\paragraph*{Strength parameter and its general definition.}
The physical measure of the transition strength is the generalized trace anomaly of the total stress-energy tensor across the wall~\cite{Barni:2025mud},
\begin{equation}
\label{eq:alpha_general}
\alpha_\vartheta \;\equiv\; \frac{D\vartheta}{3w_+}\ , \qquad D\vartheta=De-\frac{\delta e}{\delta p}Dp=D\tilde T^\mu_{\ \mu}
\end{equation}
where $Df=f_+(T_+)-f_-(T_+), \ \delta f=f_-(T_+)-f_-(T_-)$, signal the generalisation of the trace anomaly and where $w_+\equiv e_++p_+$ is the enthalpy \emph{ahead} of the wall.\footnote{This is the quantity that controls the sign of the generalized pseudo-trace introduced in Ref.~\cite{Barni:2025mud} and neatly distinguishes \emph{inverse} ($\alpha_\vartheta<0$) from \emph{direct} ($\alpha_\vartheta>0$) transitions.} Using the Bag EoS, \cref{eq:bag_eos}, one finds that this definition boils down to the usual strength of the transition defined in the literature,
\begin{equation}
\alpha_\vartheta \;=\; \frac{4\,\Delta V}{3\,w_+} \;=\; \frac{\Delta V}{\rho_{\rm rad}(T_+)} \;\equiv\; \alpha_+\,.
\end{equation}
Therefore, in the Bag model the general definition, \cref{eq:alpha_general} coincides \emph{exactly} with the familiar $\alpha_+$ used in hydrodynamic matching. In particular, the sign of $\alpha_\vartheta$ reduces to the sign of $\alpha_+$, see Ref.~\cite{Barni:2025mud}, so that within the Bag model direct PTs correspond to $\alpha_+>0$, and inverse PTs to $\alpha_+<0$.

Eliminating $p$ and $w$ in \cref{eq:junctionAB} yields the standard relation between the fluid velocities $v_\pm$ in the wall rest frame,
\begin{equation}
\label{eq:vplus_vminus_relation}
v_+(v_-, \alpha_+) = \frac{1}{1+\alpha_+} \bigg[\bigg(\frac{v_-}{2}+ \frac{1}{6v_-}\bigg) \pm \sqrt{\bigg(\frac{v_-}{2}+ \frac{1}{6v_-}\bigg)^2 + \alpha_+^2 +\frac{2}{3}\alpha_+- \frac{1}{3}}\bigg] .
\end{equation}
The two signs correspond to the familiar branches shown in the \emph{left panel} of \cref{fig:hydro_solution_space}, where the color encodes the normalized strength\footnote{This normalization is convenient to have a direct indication of the fraction of the energy budget converted into fluid kinetic energy.}, $\tilde{\alpha}_+=\alpha_+ /(1+|\alpha_+|))\in[-1,1]$; negative values~(red) indicate inverse transitions, positive values~(blue) direct ones.

It is also useful to define the strength at the nucleation temperature,
\begin{equation}
\alpha_N \equiv \frac{\Delta V}{\rho_{\rm rad}(T_N)} \, .
\end{equation}
The \emph{right panel} of \cref{fig:hydro_solution_space} summarizes the solution space in the $(\alpha,\xi_w)$ plane, where $\xi_w$ denotes the wall speed in the frame of the center of the bubble. When classifying by $\alpha_N$ (solid curves), each point $(\alpha_N,\xi_w)$ maps to a \emph{unique} hydrodynamic solution. In contrast, when classifying by $\alpha_+$ (dashed curves), there exists a meshed blue region in which the same pair $(\alpha_+,\xi_w)$ corresponds to two distinct solutions --- a \emph{detonation} or a \emph{hybrid} --- reflecting the multi-valued nature of the branches in \cref{eq:vplus_vminus_relation}. The legend specifies the color scheme, and the shaded gray area indicates kinematically forbidden configurations.

The criterion for deciding whether the fluid velocity in the frame of the center of the bubble is negative or positive --- i.e.\ whether the transition is inverse or direct --- is governed by the sign of the generalized pseudo-trace introduced in Ref.~\cite{Barni:2025mud}. As just discussed, for the Bag EoS this sign is precisely the sign of $\alpha_+$, thereby providing a simple, operational discriminator that connects directly to the matching conditions and to the branch structure visible in the \emph{left panel} of \cref{fig:hydro_solution_space}.

\subsection{Hydrodynamics of inverse phase transitions}

\begin{figure}[t]
    \centering
    \includegraphics[width=0.32\textwidth]{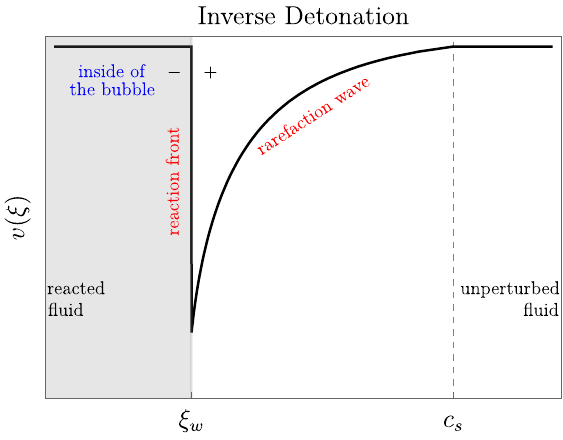}
    \includegraphics[width=0.32\textwidth]{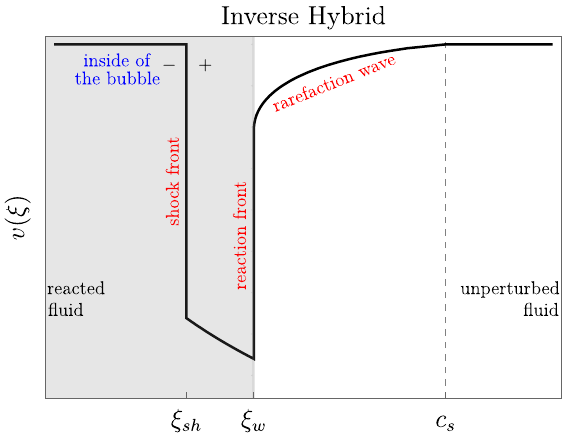}
    \includegraphics[width=0.32\textwidth]{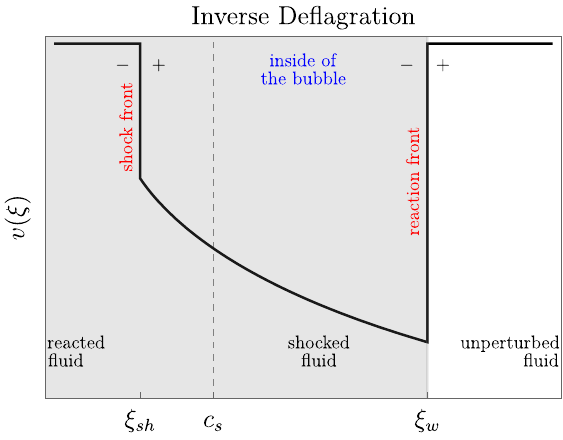}
    \caption{{\bfseries Hydrodynamic profile of the inverse phase transition.} Left Panel: inverse detonations. Middle Panel: inverse hybrid. Right Panel: inverse deflagration.}
    \label{fig:schematic}
\end{figure}

We now turn to the inverse solutions, defined by a negative bulk fluid velocity in the bubble-center frame: rather than being pushed outward, the surrounding plasma is sucked into the expanding bubble. 
We work with the self-similar variable $\xi=r/t$ and the steady solutions of the matching equations derived above, following the detailed analyses in Refs.~\cite{Barni:2024lkj, Bea:2024bxu}. Inverse expansions admit the same broad classes as in the direct case --- weak and Chapman-Jouguet (CJ) detonations, weak and CJ deflagrations, and hybrids --- but with the flow directed inward; schematic profiles are displayed in the left, middle, and right panels of \cref{fig:schematic}.

An inverse detonation is realized by joining a reaction front (namely, the bubble wall) located at $\xi_w$ with $\xi_w=v_-<c_{s,-}$, where we $c_{s,\pm}$ is the sound velocity in the plasma on each side of the wall, to a rarefaction wave ahead of the wall. Immediately in front of the wall the plasma velocity in the plasma frame is
\[
v(\xi_w^+)=\mu(v_-,v_+)=\frac{v_--v_+}{1-v_-v_+}\,,
\]
with $v_-<v_+$, which is always negative in the detonation branch: the fluid streams toward the wall even though the bubble expands because $\xi>0$. Across the rarefaction, from $\xi=c_{s,+}$ up to $\xi=\xi_w$, both the pressure and the speed decrease monotonically and relax to $v\to 0$ at the sonic point $\xi=c_{s,+}$; the corresponding profiles are shown in the left panel of \cref{fig:schematic}. The CJ inverse detonation is obtained when the post-front flow is exactly sonic on the `$-$' side, $v_-=c_{s,-}$, so that the rarefaction attaches to the reaction front.

Starting from $v_+ < c_s$ for a weak inverse detonation, the transition to a \emph{forbidden} strong inverse detonation occurs when $v_+ > c_s$, so that the fastest allowed inverse detonation is given by the limit $v_+ \to c_s$. To find the velocity of the \emph{fastest inverse detonation}, the inverse Jouguet velocity $v^{\rm inv}_J$ we can then set $v_- = v^{\rm inv}_J$ and $v_+ = c_s$ to obtain
\begin{equation} 
\label{eq: v- in term of vp_bis}
c_s = \frac{1}{1-|\alpha_+|} \bigg[\bigg(\frac{v^{\rm inv}_J}{2}+ \frac{1}{6v^{\rm inv}_J}\bigg) -\sqrt{\bigg(\frac{v^{\rm inv}_J}{2}+ \frac{1}{6v^{\rm inv}_J}\bigg)^2 + \alpha_+^2 -\frac{2}{3}|\alpha_+|- \frac{1}{3}}\bigg],
\end{equation}
which implicitly gives $v_J^{\rm inv}$ as a function of $|\alpha_+|$.

Inverse deflagrations arise when the reaction front satisfies $v_+=\xi_w>\,v_{J,\rm inv}$ and $v_->c_{s,-}$; the fluid ahead of the wall is then processed by a compression wave which steepens into a shock at $\xi=\xi_\mathrm{sh}>\xi_w$. Far ahead of the shock, one has $v\to 0$, while across the compression both $p$ and $|v|$ increase toward the wall in accordance with the Rankine-Hugoniot conditions. The CJ inverse deflagration corresponds to the limiting case where the flow immediately ahead of the front is sonic on the `$+$' side, $v_+=c_{s,+}$, and the compression wave degenerates into a sonic attachment; a schematic example is given in the right panel of \cref{fig:schematic}.

Between these regimes, for wall speeds in the interval $c_{s,+}>\xi_w>v_{J,\rm inv}$, the steady state is an inverse hybrid. The structure consists of a rarefaction wave glued to a detonation-like reaction front, followed by a compression wave and then a terminal shock that brings the fluid to rest far in front of the bubble. The slowest admissible inverse Jouguet speed $v_{J,\rm inv}$ bounds from above the domain of inverse hybrids; choosing this minimal value yields the largest possible hybrid window,
\begin{equation}
c_{s,-}^2 < \xi_w < c_{s,+}\,,
\end{equation}
beyond which either the detonation or the deflagration branch is selected and the composite construction ceases to satisfy the matching and causality constraints. By definition,\footnote{First introduced in Ref.~\cite{Barni:2024lkj}, $v_{J,\rm inv}$ marks the transition between inverse detonation and inverse hybrid regimes and can be viewed as the wall speed of the fastest inverse detonation (or, equivalently, the slowest inverse hybrid).} $v_{J,\rm inv}$ is the boundary value at which the rarefaction and compression attachments become sonic and the hybrid solution bifurcates into a single-wave configuration.

The three families just described --- inverse detonations, inverse hybrids, and inverse deflagrations --- provide the complete set of steady inverse solutions consistent with the matching conditions, and their qualitative features are summarized in \cref{fig:schematic}. We stress that from now on we will consider the case $c_{s,-}=c_{s,+}=c_s=1/\sqrt{3}$.

After introducing these necessary preliminaries (see Ref.~\cite{Barni:2024lkj} for further details), we now turn to the inverse regime and solve for the bubble-wall velocity. Under the LTE assumption, we will obtain a prediction for the steady-state velocity, which is a crucial ingredient for computing the gravitational-wave spectrum. This will be the subject of the next section.

\section{The velocity in the LTE regime}
\label{sec:velocity}

\noindent In principle, the bubble wall velocity is not fixed by hydrodynamical considerations alone, and requires input from the microphysics of the plasma. This can be seen by counting the number of quantities that describe the transition, four, $T_+, T_-, v_+, v_-$, compared to the two conservation equations of the stress-energy tensor, \cref{eq:matching-original}. Hence, even for a given nucleation temperature (which is determined by the nucleation rate), one degree of freedom remains undetermined, which we typically take to be the wall velocity $\xi_w$. Determining $\xi_w$ from first principles requires a detailed study of the microscopic friction exerted by the plasma on the expanding wall.

A widely used approach to estimate the wall velocity assumes that the system remains in \emph{local thermal equilibrium} (LTE), where the distribution functions of plasma species are close to their equilibrium form, and deviations are parametrically small. Within this approximation, the friction can be computed by expanding around equilibrium and solving linearized Boltzmann equations \cite{Ai:2021kak,Ai:2023see,Balaji:2020yrx,Krajewski:2024gma,Branchina:2025jou}. While this framework has proven useful for obtaining semi-analytic estimates of the wall dynamics, it is important to stress that LTE is only an approximation, and in fact, it is likely to fail in realistic settings. Recent studies beyond LTE have shown that out-of-equilibrium effects can substantially alter the predicted velocity, with differences up to $\sim\SIrange{50}{60}{\%}$ compared to the LTE estimate \cite{Krajewski:2024zxg,Branchina:2025adj,Ai:2025bjw,Ekstedt:2024fyq}. 

Nevertheless, for exploratory purposes, we will adopt the standard LTE treatment, which allows one to capture the qualitative behavior of the wall and obtain a reasonable order-of-magnitude estimate of its velocity. This is particularly useful for mapping the parameter space of novel scenarios, such as inverse phase transitions \cite{Barni:2024lkj,Barni:2025mud}, where a complete out-of-equilibrium computation is not yet available.

\subsection{LTE framework}

\noindent In the limit in which one can assume that the plasma inside the bubble wall remains close enough to equilibrium~\cite{Balaji:2020yrx, Ai:2021kak, Ai:2023see,Ai:2024shx}, a further matching condition permits solving for the velocity. At the intuitive level, if the collisions among particles are efficient enough inside the wall, the distributions of particles remain close to an equilibrium one, parameterized only by the temperature. The evolution of thermodynamical quantities inside the wall is thus well captured by a hydrodynamical \emph{continuous} wave. As it is well known, continuous waves conserve the entropy current $s u^\mu$, such that $
\partial_\mu(s u^\mu) = 0 
$\footnote{In fact, for the phase boundary, the presence of the scalar profile, if the wall is thick enough, can help prevent the development of the discontinuity in hydrodynamical quantities and local thermal equilibrium is enforced (see however, for a recent study of the entropy production inside the bubble wall~\cite{Eriksson:2025owh}). Notice that this cannot be the case for the genuine shock propagating in front of the bubble wave for deflagrations.  In this case, the plasma unavoidably develops a genuine discontinuous wave at the position of the shock wave, across which $\partial_\mu(s u^\mu) = 0$ cannot be fulfilled. This point was the source of some confusion recently, for example, in Ref.~\cite{Yuwen:2024hme}.}.  The immediate consequence of assuming local thermal equilibrium is that there is one more matching condition, which thus fixes the last parameter left unconstrained by hydrodynamics, the wall velocity. 
 
Going back to the system of equations in \cref{eq:junctionAB}, one can observe that a non-vanishing value for the term $\int_{-\delta}^\delta \text{d} z\, (\partial_z\phi)\, \ddot{\phi}(0)|_{\tau}$ would mean that the wall is accelerating (or decelerating), consequently, searching for stationary wall, we set it to zero and study the rest of the system. This system has a stationary solution if there exists a pressure from the plasma which balances the pressure from the vacuum
\begin{equation}
\underbrace{\mathcal{P}_{\rm driving}}_{\text{pressure from vacuum}} - \underbrace{\mathcal{P}_{\rm LTE}}_{\text{pressure from the plasma}}  = 0\, ,
\end{equation} 
that is to say, if the pressure from the plasma in equilibrium can compensate the pressure from the vacuum potential. We defined $\mathcal{P}_{\rm driving} \equiv \frac{3}{4} w_+ \alpha_+$. Notice that the sign of each contribution depends on the type of transition: for a direct PT, the pressure from the vacuum is positive (the vacuum pushes the wall) and the friction from the plasma negative (it resists the expansion), while the opposite is realised for the inverse case.

 It is clear that the pressure from the plasma is a strong function of the phase boundary velocity $\xi_w$. Using the system of equations in \cref{eq:junctionAB}, one immediately observes that 
\begin{align}
\label{eq:P-LTE-def}
    \mathcal{P}_{\rm LTE}= -\Delta V_T+\overline{\mathcal{P}}_{\rm LTE},
\end{align}
where
\begin{align}
\label{eq:Pbar-LTE-def}
    \overline{\mathcal{P}}_{\rm LTE}\equiv \Delta  \{w \gamma^2 v^2\}  =  \Delta \{ (\gamma^2-1)Ts\}
\end{align}
with $s=w/T$ being the entropy density. Using \cref{eq:conditionA}, one can rewrite it as
\begin{align}
\label{eq:LTE-pressure2}
    \overline{\mathcal{P}}_{\rm LTE} (w_+, v_+, v_-)= w_+ \gamma_+^2 v_+ (v_+-v_-).
\end{align}
This pressure can have both sign depending of the difference $(v_+-v_-)$, so that  the quantity $\overline{\mathcal{P}}_{\rm LTE} $ can be either pushing or slowing the bubble wall. 
On the other hand, the difference of thermal potential is given by
\begin{align}
    -\Delta V_T = \frac{w_+}{4}\bigg(1 - \frac{a_-}{a_+} \bigg(\frac{T_-}{T_+}\bigg)^4 \bigg) \, .
\end{align}
 This result is obtained in the following way. In the high temperature regime $T\gg m$, the thermal potential is given by 
\begin{align}
\label{eq:VTT}
     V_T(T) = \frac{w (T)}{4} + T^4\mathcal{O}\bigg(\frac{m^2}{T^2}\bigg) ,
\end{align}
while the contribution from massive particles with $m\gg T$ is exponentially suppressed and scales like $e^{-m/T}$. The potential in the Bag model is obtained by neglecting the subleading thermal corrections in \cref{eq:VTT}, thus considering only the light species in each phase.
In LTE, there is an additional matching condition due to the entropy current conservation~\cite{Ai:2021kak}
\begin{equation} 
\label{eq:LTE-matching_1}
   \partial_\mu (s u^\mu) = 0\quad\Rightarrow\quad  s_+ \gamma_+ v_+ =s_- \gamma_- v_- \quad \text{(matching condition from entropy current conservation)}.
\end{equation}
and we can recast the ratio of temperatures in the ratio of velocities
\begin{align}
 \label{eq:LTE-matching}
\frac{T_+}{T_-}=\frac{\gamma_-}{\gamma_+}\,.
\end{align}

\Cref{eq:LTE-matching} allows to eliminate the temperature in the different contributions to the pressure. The thermal part of the difference in potential becomes 
\begin{align}
    -\Delta V_T (w_+, v_+, v_-)= \frac{w_+}{4}\bigg(1 - \Psi \bigg(\frac{\gamma_+}{\gamma_-}\bigg)^4 \bigg) ,
\end{align}
where we have defined $\Psi \equiv a_-/a_+$. We observe that within the LTE approximation, both $\Delta V_T$ and $\overline P_{\rm LTE}$ are only function of the three variables $(w_+, v_+, v_-)$. We expect that for a \emph{direct} phase transition, some of the particles gain mass and become non-relativistic, so $a_- < a_+$, or $\Psi <1$ typically. On the other hand, in the case of an inverse phase transition, some particles typically lose their mass while others remain massless and we have $a_- > a_+$, or $\Psi >1$. 
Notice, however, that the sign of $\Delta V_T$ is not fixed since the ratio $\gamma_+/\gamma_-$ could be either larger or smaller than one. 

Finally, the complete pressure on the bubble wall in the LTE regime  is given by
\begin{equation}
\label{eq:pres}
\mathcal{P}^{}_{\rm bubble}(v_+) = \mathcal{P}_{\rm LTE}-\mathcal{P}_{\rm driving}  = \frac{w_+}{4} \bigg(4 \gamma_+^2 v_+(v_+-v_-) + 1 - \Psi\bigg(\frac{\gamma_+}{\gamma_-} \bigg)^4 - 3 \alpha_+(\xi_w)\bigg) \, . \qquad \qquad \text{(LTE pressure)}
\end{equation}

With this equation at hand, we can study the pressure for the direct and the inverse phase transitions.
First of all, we can study this relation in the case of a just-nucleated bubble. The bubble is taken to be nucleated at rest with a vanishing wall velocity, $\xi_w \to 0$, and correspondingly $v_+ = v_- \to 0$, consequently we have that  
\begin{equation}
\mathcal{P}^{}_{\rm bubble}(v_+)\bigg|_{\xi_w = v_+ = v_- \to 0} \to \frac{w_+}{4} \bigg( 1 - \Psi - 3 \alpha_N\bigg) \, . 
\end{equation}

If this pressure is positive, then the bubble cannot expand from the start. This can be understood from the fact that in the region of the parameter space with 
\begin{equation}
\label{eq:cond_exp}
\alpha_N < \frac{1-\Psi}{3}, \qquad \qquad  \text{(Nucleation thermodynamically disfavored)} \, ,
\end{equation}
the bubble nucleation itself is thermodynamically disfavored~\cite{Espinosa:2010hh}. This condition imposes a lower bound on $\alpha_N$ for a given $\Psi \neq 1$, which is displayed in \cref{Fig:velocity} by the ``no expansion'' line. Even if this condition is the same for the direct and inverse phase transitions, its interpretation significantly differs.
For direct phase transitions, $a_- < a_+$ and thus $\Psi < 1$, $\alpha_N$ is positive and encodes the outward pressure from the vacuum while $\Psi$ encodes the pressure from the plasma over the bubble. The condition in \cref{eq:cond_exp} is the condition that the vacuum pressure is too small to balance the plasma inward pressure. On the other hand, for inverse transitions where $a_- > a_+$ and thus $\Psi > 1$, $\alpha_N$ is negative and the large pressure from the vacuum is  preventing the expansion while the plasma pressure favors the expansion. \cref{eq:cond_exp} now encodes the fact that the inward vacuum contribution dominates.

\subsection{LTE pressure for direct phase transitions}

\begin{figure}[t]
    \centering
    \includegraphics[width=\linewidth]{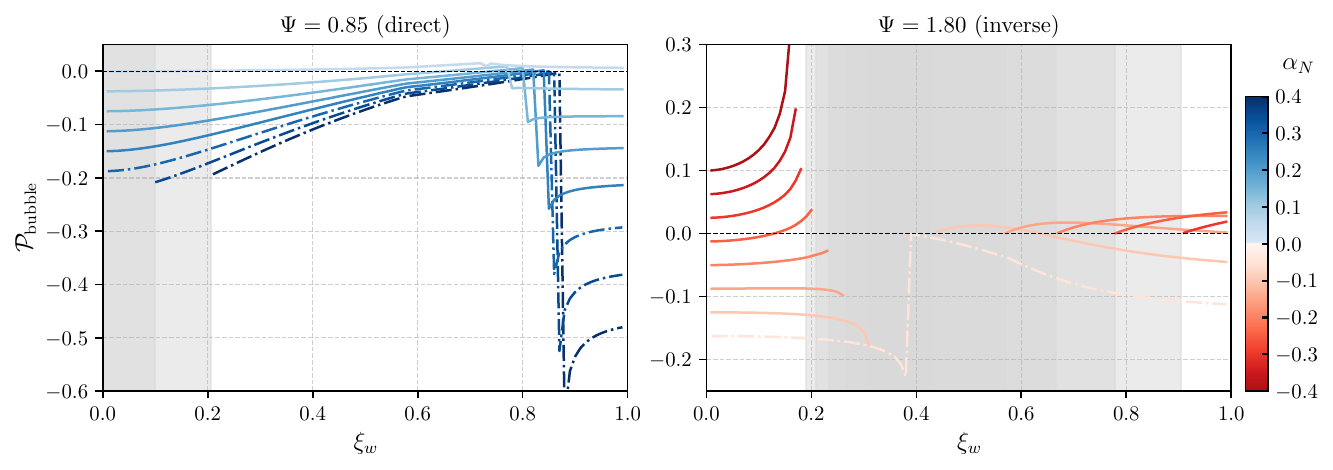}
    \caption{{\bfseries\boldmath Profile pressure for different values of $\Psi, \alpha_N$.} Left is for direct PT with $\Psi=0.85$, while right is for inverse PT with $\Psi=1.8$. Dot-dashed lines correspond to runaway solutions, where the friction never vanishes and remains negative throughout the entire wall evolution.} 
    \label{Fig:velocity_curves}
\end{figure}

The pressure on the bubble wall assuming (local) equilibrium distributions has been abundantly explored in the context of \emph{direct} phase transitions \cite{Balaji:2020yrx, Ai:2021kak, Ai:2023see,Ai:2024shx}. On the left panel of \cref{Fig:velocity_curves} we display the pressure from \cref{eq:pres} in the context of a direct PT. We observe that the pressure always displays a peak at the Jouguet velocity\footnote{Notice that it has been recently suggested that the dynamical evolution of the bubble might prevent the peak from stopping the acceleration of the wall~\cite{Krajewski:2024zxg, Krajewski:2024gma}.}.  
We can also observe that the pressure is always positive for $\alpha_N < (1-\Psi)/3$, reflecting the fact that the expansion of such bubbles is not thermodynamically favorable.

\begin{figure}[ht]
    \centering
    \includegraphics[scale=0.8]{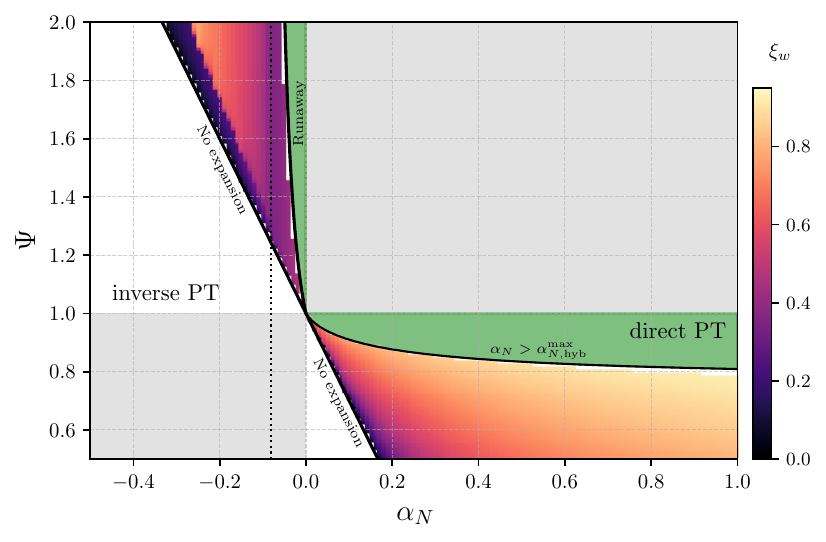}
   \caption{%
{\bfseries\boldmath Phase diagram of the LTE wall velocity in the $(\alpha_N,\Psi)$ plane.} 
Coloured squares show the numerical wall speed.
The solid black boundary is the full $\mathcal{P}_{\mathrm{LTE}}=0$ contour:
its straight left segment is the static-wall limit
$\mathcal{P}_{\mathrm{LTE}}(\xi_w=0)=0$ (a newly nucleated bubble does not expand).
In the direct region, the curved branch marks the onset of runaway,
equivalently $\mathcal{P}_{\mathrm{LTE}}(\xi_w=\xi_J)<0$;
in the inverse region, it is set by the slowest hybrid just past the inverse
Jouguet speed, $\mathcal{P}_{\mathrm{LTE}}(\xi_w=\xi_{J,\mathrm{inv}}^+)<0$.
The green band highlights parameters satisfying the runaway criterion.
The dotted vertical line at $\alpha_N\simeq-0.082$ indicates
the point at which the kinematically forbidden region opens up, and the slowest inverse hybrid solution is always a stable solution.}
    \label{Fig:velocity}
\end{figure}

In \cref{Fig:velocity}, we show the terminal velocity of the bubble as computed within the LTE regime by solving $\mathcal{P}_{\mathrm{LTE}}(\xi^{\rm LTE}_w=\xi)=0$. 
The lower right part of the plot concerns direct phase transitions with positive $\alpha_N$. The green region is the region that exhibits runaway solutions if one considers only the LTE source of pressure. The boundary can be computed following the methods presented in Refs.~\cite{Ai:2023see, Ai:2024shx}. 

Looking back at \cref{Fig:velocity_curves}, one observes that in multiple cases, the equation $\mathcal{P}_{\mathrm{bubble}}(\xi^{\rm LTE}_w=\xi)=0$ has several solutions. These solutions are not guaranteed to be stable: the condition for the stability of a solution is given by
\begin{equation}
\label{eq:stability}
    \frac{\text{d}}{\text{d}\xi_w} \mathcal{P}^{\rm }_{\rm bubble}(\xi_w^{\rm LTE}) >0 \, .
\end{equation}
If such a condition is not satisfied, any small increase of the wall velocity around the LTE solution $\xi_w^{\rm LTE}$ would imply an acceleration that would take the system away from the solution itself, as is typical of instabilities. 

\subsection{The LTE pressure for inverse phase transitions}

We now turn to the study of the pressure on the bubble wall in the context of inverse phase transitions within LTE. On the right panel of \cref{Fig:velocity_curves}, we present the pressure curves for the inverse phase transitions and on \cref{Fig:velocity}, we present the terminal velocity of the wall using the criterion $\mathcal{P}_{\mathrm{bubble}}(\xi^{\rm bubble}_w=\xi)=0$.  First of all, as in the previous discussion, one can  determine
the leftmost boundary of the velocity scan  by the requirement that the static‐limit pressure vanishes,
\begin{equation}
\mathcal{P}_{\rm bubble}(\xi_w=0)=0 \qquad \Rightarrow\qquad \mathcal{P}^{}_{\rm bubble}\Big|_{\xi_w = v_+ = v_- \to 0} \to \frac{w_+}{4} \bigg( 1 - \Psi - 3 \alpha_N\bigg) \, . 
\end{equation}
We conclude similarly that all bubbles with $\alpha_N < \frac{(1-\Psi)}{3} \, ,$
cannot expand. 

\paragraph{The vanishing pressure of the slowest hybrid}

In Ref.~\cite{Barni:2024lkj}, it was noticed that for $\alpha_N < \alpha_{N,\rm slowest}$, a kinematically forbidden gap appears around the inverse Jouguet speed. 
When such a region of forbidden wall velocity $\xi_w$ exists, then at the slower boundary of the hybrid solutions, the self-similar solution is such that:
\begin{equation}
   \text{(Slowest hybrid):} \qquad \qquad  v_+ = c_s, \qquad v_- \to  1, \qquad \gamma_- \to \infty \,.
\end{equation} 
We emphasize that this solution might be artificial, as it involves luminal fluid velocities, however it should be thought as a well behaved limit.

Let us study the pressure on this specific solution by inspecting \cref{eq:pres}. The fact that $\gamma_- \to \infty$ implies that the term containing $\Psi$ in the pressure drops out, making the pressure \cref{eq:pres} $\Psi-$independent. On the other hand, let us also remind that the value of $\alpha_+$ is totally fixed to be 
\begin{equation}
\text{(Slowest hybrid):} \qquad \qquad  v_+\big|_{\alpha_+= \alpha_{+,\rm slowest}} =  c_s = \frac{2}{3}\bigg(\frac{1}{1-|\alpha_+|} + \sqrt{\frac{1}{3} + \alpha_+^2 - \frac{2}{3}\alpha_+}  \bigg) \bigg|_{\alpha_+= \alpha_{+,\rm slowest}}\, . 
\end{equation}

Solving for $\alpha_+$, we obtain $\alpha_{+,\rm slowest} = 1 - \frac{2}{\sqrt{3}} \approx - 0.154$. First of all this observation allows us to recover the value of $\alpha_N$ for which the slowest hybrid solution merges with the inverse detonation branch, i.e. the $\alpha_{N,\rm slowest}$. Constructing numerically the profile of this slowest hybrid allows to determine its corresponding $\alpha_{N}$:
\[
\alpha_{+,\rm slowest} = 1 - \frac{2}{\sqrt{3}}
\quad\Longrightarrow\quad
\alpha_{N,\rm slowest}\approx -0.082\,.
\]
 We draw this boundary in dashed black on \cref{Fig:velocity}, all the phase transitions on the left of this boundary display a gap between the hybrid and the detonation branch.

Secondly, using $\alpha_{+,\rm slowest} = 1 - \frac{2}{\sqrt{3}}$ in the expression for the pressure in \cref{eq:pres}, we observe that the LTE pressure rigorously vanishes in this limit
\begin{equation}
\mathcal{P}^{\text{slowest hybrid}}_{\rm bubble} \to 0 \, . 
\end{equation} 

This trend, which was argued on analytical arguments, is confirmed numerically on the right panel of \cref{Fig:velocity_curves}. We directly observe that the slowest hybrid, where the second part of the curve begins, is exactly zero. 

On the other hand, if $\alpha_N > \alpha_{N,\rm slowest} \approx -0.082$, there is no forbidden region for the velocity, and consequently the slowest hybrid does not exhibit $v_- \to 1$. 

\paragraph{Runaway region:}

On \cref{Fig:velocity}, we delineated via a black solid line the boundary between the terminal velocity and the runaway regions. Let us get a taste of how we can deduce this boundary with analytical methods. Starting from the inverse hybrid‐wave matching conditions with upstream velocity set to the speed of sound,
\[
v_+ = c_s,\qquad v_- = v_-\bigl(c_s,\alpha_+\bigr)\,,
\]
we notice that this solution, in the absence of a gap between the detonation and hybrid branch, i.e. $\alpha_N > \alpha_{N,\rm slowest}$, maximises the LTE pressure (see for example the pale orange continuous curve of the left panel of \cref{Fig:velocity_curves}). If this maximal pressure is negative (positive), then the bubble wall is expected to runaway (reach a terminal velocity). The boundary between these two behaviours should be determined by a vanishing maximal pressure: 
consequently imposing \(\mathcal{P}_{\rm bubble}(\xi_w= \xi_{\text{slowest  hybrid}})=0\), one can extract the parameter \(\Psi_{\rm boundary}\) 
\begin{equation}\label{eq:b_inverse_hybrid}
\Psi_{\rm boundary} \;=\;
\frac{4\bigl(1 - \alpha_+ - 2\sqrt{\alpha_+(\alpha_+-2)}\bigr)}
     {9\Bigl[\,1 - \tfrac{1}{3}\bigl(1 - \alpha_+\,\sqrt{\alpha_+(\alpha_+-2)}\bigr)^2\Bigr]^2}\,.
\end{equation}
As such, this boundary is expressed in terms of $\alpha_+$. To express it in terms of $\alpha_N$, we then rescale
\(
\alpha_+ \to 
\alpha_N \,\frac{\alpha_{+,\rm slowest}}{\alpha_{N,\rm slowest}} \, ,
\)
and substitute into \cref{eq:b_inverse_hybrid} to obtain the black curve delineating the transition between LTE‐velocity solutions and true runaway.

On the other hand, for \(\alpha_N < \alpha_{N,\rm slowest}\), a kinematically forbidden gap appears around the inverse Jouguet speed, and in this case the slowest inverse hybrid always satisfies \(\mathcal{P}_{\rm bubble}=0\), and an LTE solution remains.  Thus \(\alpha_{N,\rm slowest}\) marks the onset of genuine runaway behavior.

Let us notice here that the LTE stability of our solution is formally the same as in \cref{eq:stability}. In this case the driving force is given by $\mathcal{P}_{\rm bubble}$, which needs to compensate a potentially large vacuum pressure that would here prevent the bubble to expand. One can similarly derive that:
\begin{equation}
    \frac{\text{d}}{\text{d}\xi_w} \mathcal{P}^{\rm }_{\rm bubble}(\xi_w^{\rm LTE}) >0.
\end{equation}
One may also conclude that points with negative $\mathcal{P}^{\rm }_{\rm bubble} < 0$ are in principle runaways. 

\paragraph{Runaway and LTE}
To conclude with a word of caution: 
the analysis presented in this section is only per se valid if one assumes the validity of the local thermal equilibrium approximation inside the wall. Specifically, in the runaway regime, the LTE approximation is expected to break down badly. In this case, the ballistic (or kick) picture is more adequate, as discussed in Ref.~\cite{Barni:2024lkj} and several other sources of pressure on the bubble wall can enter the picture, including the particles gaining mass\cite{Bodeker:2009qy, Ekstedt:2024fyq} and the emission of particles for the direct\cite{Bodeker:2017cim,Gouttenoire:2021kjv, Azatov:2020ufh,Ai:2023suz, Ai:2025bjw} and inverse\cite{Barni:2024lkj, Barni:2025mud} transitions. Such new contributions can often halt the acceleration of the bubble wall and impose a terminal velocity. Consequently, the region which we call ``runaway'' in this paper should be considered with care and would necessitate a specific model-dependent study.  

\section{Sound-shell model in an expanding Universe}
\label{sec:sound_shell}

In this section, we compute the gravitational-wave signal from first-order phase transitions with the explicit goal of disentangling \emph{direct} from \emph{inverse} dynamics. Our strategy is to model the acoustic stage within the sound-shell framework, in which the velocity field produced by the expanding bubbles is described as a superposition of spherically symmetric shells whose overlap sources tensor perturbations on a Friedmann background.
Over the past years the framework has been thoroughly developed and applied to phase transitions of weak to intermediate strength. Starting from its original formulation and first quantitative predictions~\cite{Hindmarsh:2016lnk,Hindmarsh:2019phv}, it has been used extensively in observational studies~\cite{Gowling:2021gcy,Boileau:2022ter,Gowling:2022pzb,Guo:2024gmu,Tian:2025zlo}, and subsequently generalised to account for the role of the cosmological expansion in shaping the build-up and time coherence of the source~\cite{Guo:2020grp} as well as for non-standard equations of state~\cite{Giombi:2024kju}.
Here we adopt the formulation that clarifies the infrared momentum dependence and the emergence of intermediate spectral slopes~\cite{RoperPol:2023dzg}, which we implement to track how the hydrodynamic differences between direct and inverse transitions propagate into the shape and normalization of the spectrum.

Concretely, we evolve tensor modes on a radiation-dominated background using the linear response of the metric to the transverse-traceless part of the fluid stress. The sound shells encode the single-bubble profiles fixed by hydrodynamics --- compression and rarefaction waves ahead/behind of the wall --- and their superposition determines the acoustic power available to source gravitational waves. The finite lifetime of the acoustic stage and the Hubble expansion enter through the time kernel that weights the unequal-time correlator of the anisotropic stress, so that long-lived sources in Hubble units accumulate more power and imprint a sharper shoulder near the peak, whereas short-lived sources remain comparatively suppressed. Within this setup, the parameters that control the signal are the wall speed and strength extracted from the matching conditions, the kinetic energy fraction carried by the flow, the mean bubble separation that sets the characteristic wavenumber, and the effective duration of the acoustic phase.

The key advantage of this framework for our purposes is its transparency: once the hydrodynamic branch is fixed, the differences between outflowing (direct) and inflowing (inverse) solutions map cleanly onto the sound-shell geometry and thus onto the gravitational-wave spectrum. In what follows, we use the implementation of the SSM as in Ref.~\cite{RoperPol:2023dzg}.
We will first lay out the governing equations and source decomposition, then present the kinetic and anisotropic-stress spectra that serve as inputs, and finally assemble the gravitational-wave spectra highlighting the features that distinguish direct from inverse transitions in both the infrared rise and the near-peak structure.
We here closely follow the derivation in Refs.~\cite{Hindmarsh:2019phv,RoperPol:2023dzg}.
The reader familiar with the sound shell model might directly skip this section and go to \cref{sec:results}.

\subsection{Tensor equation, source statistics, and the spectrum}

To set the stage, we work on a spatially flat Friedmann background with conformal time \(\tau\) and scale factor~\(a(\tau)\), \(ds^2 = a^2(\tau)\big[-d\tau^2 +\big(\delta_{ij}+h_{ij}(\tau,\mathbf{x})\big)\, dx_i  dx_j\big]\). Gravitational-wave tensor perturbations \(h_{ij}\) are defined as transverse and traceless metric fluctuations, \(\partial_i h_{ij}=0\) and \(h_{ii}=0\), and obey the linearized Einstein equation
\begin{equation}\begin{aligned}
    \Box h_{ij}(\tau,\mathbf{x}) =  \left(\partial_\tau^2 + 2\mathcal{H}(\tau)\,\partial_\tau - \nabla^2\right) h_{ij}(\tau,\mathbf{x}) &= 16\pi G\,a^2(\tau)\, \bar{\rho}(\tau)\,\Pi_{ij}(\tau,\mathbf{x})\,,\\
    \label{eq:EoM-SSM}
    \Longrightarrow\qquad \left(\partial_\tau^2 + k^2 - \frac{a''(\tau)}{a(\tau)}\right) a(\tau) h_{ij}(\tau,\mathbf{k})
    &= 16\pi G\,a^3(\tau)\, \bar{\rho}(\tau)\,\Pi_{ij}(\tau,\mathbf{k})\,,
\end{aligned}\end{equation}
where prime denotes \(\partial_\tau\), \(\mathcal{H}\equiv \frac{a'}{a}\) is the conformal Hubble rate, and
\(\Pi_{ij} = \Lambda_{ij,mn} T_{mn}/\bar{\rho}\) is the dimensionless transverse-traceless (TT) part of the source anisotropic stress tensor (constructed from the fluid velocity field) normalized to its mean energy density \(\bar{\rho}\), with \(\Lambda_{ij,mn}\) being the transverse-traceless projector.
The second line is  obtained by going to Fourier space.

During radiation domination \(a(\tau)\propto \tau\) and thus \(a''/a=0\), so the cosmic expansion affects the signal through the explicit time dependence of the source and the finite duration of the acoustic stage rather than through an effective mass term.
The retarded Green’s function for \cref{eq:EoM-SSM} is then  \(G_k(\tau,\tau')=\frac{\sin[k(\tau-\tau')]}{k}\,\Theta(\tau-\tau')\). The solution at time \(\tau\) is therefore, using that the source red-shifts like radiation and that \(\mathcal{H}=1/\tau\) during radiation domination,
\begin{equation}
\label{eq:greens-solution}
    h_{ij}(\tau,\mathbf{k}) = \frac{16\pi G\,a^2(\tau)\,\rho(\tau)}{\mathcal{H}(\tau)}\,\int_{\tau_*}^{\tau_{\rm fin}}\!\frac{d\tau'}{\tau'}\,G_k(\tau,\tau')\,\Pi_{ij}(\tau',\mathbf{k})\,,
\end{equation}
where \(\tau_*\) is the onset of the source --- here identified with the time of bubble collisions that seed the acoustic stage --- and \(\tau_{\rm fin}\) marks its termination, when coherent sound waves decay (e.g.\ by nonlinear damping or the onset of turbulence). For later use, we introduce the active duration \(\delta\tau_{\rm fin}\equiv\tau_{\rm fin}-\tau_*\) as a control parameter of the build-up.

The gravitational-wave energy density can be expressed in terms of the tensor time derivatives. For modes well inside the horizon one may use
\begin{equation}
    \rho_\mathrm{GW}(\tau)\;=\;\frac{1}{32\pi G\,a^2}\,\big\langle h_{ij}'(\tau,\mathbf{x})h_{ij}'(\tau,\mathbf{x})\big\rangle
    \;=\;\frac{1}{32\pi G\,a^2} \int\!\frac{d^3k}{(2\pi)^3}\,\int\!\frac{d^3q}{(2\pi)^3}\,\big\langle  h_{ij}'(\tau,\mathbf{k})\, h_{ij}'(\tau,\mathbf{q})\big\rangle,
\end{equation}
where angular brackets denote an ensemble average. 
Using \cref{eq:greens-solution} and statistical homogeneity, the two-point function of the source is fully specified by the unequal-time correlator (UETC) of the anisotropic stress, which we define by
\begin{equation}
\label{eq:defUETC}
    \big\langle \Pi_{ij}(\tau_1,\mathbf{k})\,\Pi_{ij}(\tau_2,\mathbf{k}')\big\rangle
    \equiv (2\pi)^3\delta^{(3)}(\mathbf{k}+\mathbf{k}')\,\frac{2 \pi^2}{k^2} E_\Pi(\tau_1,\tau_2,k)\,.
\end{equation}
Upon substituting \cref{eq:greens-solution,eq:defUETC} into the expression for \(\rho_{\rm GW}\), one obtains a double time integral over the UETC modulated by the cosine kernel inherited from the Green’s function. 
Assuming that the source is the dominant contribution to the total energy density, so that \(\mathcal{H}^2 = 8 \pi G a^2 \bar{\rho}/3\), we obtain the fractional energy density per logarithmic interval
\begin{equation}\begin{aligned}
    \Omega_\mathrm{GW}(\tau,k) \;\equiv\; 
    \frac{1}{\bar{\rho}} \frac{\dd \rho_\mathrm{GW}}{\dd\log k} \;&=\;
    3 
    \int_{\tau_*}^{\tau_{\rm fin}}\!\frac{d\tau_1}{\tau_1}\!
    \int_{\tau_*}^{\tau_{\rm fin}}\!\frac{d\tau_2}{\tau_2}\,
    k E_\Pi(\tau_1,\tau_2,k)\,\cos\!\big[k(\tau-\tau_1)\big]\cos\!\big[k(\tau-\tau_2)\big],\\
    \label{eq:Omega-production}
    &\simeq\; \frac{3}{2}\,
    \int_{\tau_*}^{\tau_{\rm fin}}\!\frac{d\tau_1}{\tau_1}\!
    \int_{\tau_*}^{\tau_{\rm fin}}\!\frac{d\tau_2}{\tau_2}\,
    k\,E_\Pi(\tau_1,\tau_2,k)\,\cos\!\big[k(\tau_1-\tau_2)\big].
\end{aligned}\end{equation}
The product of cosines decomposes into terms with phases proportional to the \emph{difference} \(\tau_-\equiv\tau_2-\tau_1\) and to the \emph{sum} \(\tau_+\equiv\tau_1+\tau_2-2\tau_0\). In the short-wavelength regime \(k\tau_0\gg1\), the \(\tau_+\) terms oscillate rapidly over the finite interval \([\tau_*,\tau_{\rm fin}]\) and average out, while the \(\tau_-\) piece remains, yielding the expression in \cref{eq:Omega-production}.

To relate the spectrum at the end of production, \(\Omega_\mathrm{GW}^*(k) \equiv \Omega_\mathrm{GW},(\tau_\mathrm{fin},k)\) to today’s observable spectrum we redshift the energy density as radiation and account for changes in the effective number of relativistic species. Writing \(\Omega_{\rm GW}(\tau_0,k)=T_{\rm GW}\,\Omega_{\rm GW}(\tau_{\rm fin},k)\), the transfer factor can be expressed as
\begin{equation}
\label{eq:Tgw}
    T_{\rm GW}\;=\; \frac{a_*^2 \mathcal{H}_*^2}{a_0^2 \mathcal{H}_0^2}
    \;=\;\Omega_{r,0}\left(\frac{g_{*,0}}{g_*}\right)\left(\frac{g_{*s}}{g_{*s,0}}\right)^{\!4/3}
    \;\simeq\; 3.57\times 10^{-5}\,\left(\frac{100}{g_*}\right)^{\!1/3},
\end{equation}
where \(g_*\) and \(g_{*s}\) are the energy and entropy relativistic degrees of freedom at production, and \(\Omega_{r,0}\) is the present radiation density fraction. 

\subsection{Kinematics and correlators in the sound-shell model}

We describe the post-collision fluid as a linear superposition of irrotational sound waves on a spatially flat FLRW background. The tensor sector is sourced by the TT part of the stress tensor and, during radiation domination, its evolution can be written in Fourier space for the field $h_{ij}$ as in \cref{eq:greens-solution}.

\paragraph{Source from the energy-momentum tensor.}
The spatial components of the energy-momentum tensor in the non-relativistic limit are
\begin{equation}
\label{eq:ssm:Tij-full}
T_{ij}=\bar w\,u_i u_j + \bar\rho\,\delta_{ij}
+\partial_i\phi\,\partial_j\phi - \tfrac12(\partial\phi)^2\delta_{ij}\,,
\end{equation}
where $\bar w=\bar\rho+\bar p$ is the background enthalpy and $u_i$ the peculiar velocity. After TT projection, the isotropic pieces and scalar-gradient terms do not contribute, so the relevant part is the quadratic fluid term. In Fourier space, the dimensionless anisotropic stress hence becomes
\begin{equation}
\label{eq:ssm:Piij-conv}
\Pi_{ij}(\tau,\mathbf k)\ =\
\Gamma \,\int\!\frac{d^3p}{(2\pi)^3}\,\Lambda_{ij,lm}(\hat{\mathbf{k}})\,u_l(\tau,\mathbf p)\,u_m(\tau,\mathbf k-\mathbf p)\,,
\end{equation}
where $\Gamma \equiv \bar{w}/\bar{\rho} = 4/3$ is the adiabatic index of the fluid, and the TT projector is defined in Fourier space by
\begin{equation}
\label{eq:ssm:TT-def}
\Lambda_{ij,\ell m}(\hat{\mathbf k})
\equiv P_{i\ell}(\hat{\mathbf k})\,P_{jm}(\hat{\mathbf k})
-\tfrac12\,P_{ij}(\hat{\mathbf k})\,P_{\ell m}(\hat{\mathbf k})\,,
\qquad
P_{ij}(\hat{\mathbf k})\equiv \delta_{ij}-\hat k_i \hat k_j\,,
\end{equation}
so that it is symmetric in \(i\!\leftrightarrow\! j\) and \(\ell\!\leftrightarrow\! m\), transverse \(\hat k_i\,\Lambda_{ij,\ell m}=0\), traceless \(\Lambda_{ii,\ell m}=0\), and idempotent
\(\Lambda_{ij,\ell m}\Lambda_{\ell m,rs}=\Lambda_{ij,rs}\).
Inserting \cref{eq:ssm:Piij-conv} into \cref{eq:defUETC}, the anisotropic stress UETC becomes
\begin{multline}
    \label{eq:EPiu4}
    E_\Pi(\tau_1,\tau_2,k) = \frac{k^2}{2 \pi^2}\,\Gamma^2\, \int\!\frac{\dd^2 \hat{k}}{4 \pi}\int\!\frac{\dd^3 k'}{(2 \pi)^3} \int\!\frac{\dd^3 p_1}{(2 \pi)^3} \int\!\frac{\dd^3 p_2}{(2 \pi)^3} 
    \,\Lambda_{ab,ij}(\hat{\mathbf{k}})\,\Lambda_{ab,\ell m}(\hat{\mathbf{k}}) \\
   \times \left< u_i(\tau_1,\mathbf{p}_1) u_j(\tau_1,\mathbf{k}-\mathbf{p}_1) u_\ell(\tau_2,\mathbf{p}_2) u_n(\tau_2,\mathbf{k}'-\mathbf{p}_2) \right>
\end{multline}

\paragraph{Gaussian velocity field and the kinetic UETC.}
The sound-shell model treats the flow as irrotational and linear, so the velocity is purely longitudinal in Fourier space,
\(u_i(\tau,\mathbf k)=\hat k_i\,u(\tau,k)\).
Assuming Gaussian statistics, homogeneity and isotropy, the two-point function defines the \emph{kinetic} unequal-time correlator (UETC) $E_{\rm kin}$,
\begin{equation}
\label{eq:ssm:uu-UETC}
\big\langle u_i(\tau_1,\mathbf k)\,u_j(\tau_2,\mathbf k')\big\rangle
=(2\pi)^3\delta^{(3)}(\mathbf k+\mathbf k')\,\hat k_i\hat k_j\,
\frac{4 \pi^2}{k^2}\,E_{\rm kin}(\tau_1,\tau_2,k)\,,
\end{equation}
so that the root-mean-squared fluid velocity $\bar{U}_f^2$  and the kinetic energy density fraction $\Omega_K$ of the flow are given by
\begin{equation}
    \label{eq:ssm:OmegaK_and_Urms}
    \bar{U}_f^2 \;\equiv\; \langle u^2(x)\rangle = 2 \int_0^\infty E_{\rm kin}(k)\,dk
    \,,\qquad
    \Omega_K \;\equiv\; \frac{\langle T_{ii}^{uu}\rangle}{\bar{\rho}} \;=\; \Gamma\, \bar{U}_f^2 \,, 
\end{equation}
where $T_{ii}^{uu} = \bar{w} u^2$ is the kinetic energy part of the energy-momentum tensor in \cref{eq:ssm:Tij-full}.

As the velocity field is assumed to be Gaussian, Wick’s theorem factorizes the four-point function of velocities in \cref{eq:EPiu4} into the product of velocity two-point functions, 
\(
    \langle u_i u_j u_\ell u_m \rangle 
    = \langle u_i u_j\rangle \langle u_l u_m\rangle
    + \langle u_i u_\ell\rangle \langle u_j u_m\rangle
    + \langle u_i u_m\rangle \langle u_j u_\ell\rangle
\), where the first contraction vanishes upon TT projection.
The Dirac-$\delta$ distributions in from \cref{eq:ssm:uu-UETC} collapse the integrals over \(k'\) and \(p_2\), and projecting with the TT operator $\Lambda_{ij,\ell m}(\hat{\mathbf k})$ generates the angular kernel
\begin{equation}
    \label{eq:ssm:TT-kernel}
    \Lambda_{ij,\ell m}(\hat{\mathbf k})\,\hat p_i \hat{\tilde{p}}_j\hat{\tilde{p}}_\ell \hat p_m
    = \frac{p^2}{\tilde p^2}\,\frac{(1-z^2)^2}{2},
    \quad z\equiv \hat{\mathbf k}\!\cdot\!\hat{\mathbf p}\,, 
    \quad \tilde p\equiv|\mathbf k-\mathbf p|=\sqrt{k^2+p^2-2kp\,z}\,,
    \quad \hat{\tilde{p}}\equiv\frac{{\mathbf k}-{\mathbf p}}{\tilde{p}}\,.
\end{equation}
\Cref{eq:EPiu4} hence becomes
\begin{equation}
\label{eq:ssm:Epi-final}
E_\Pi(\tau_1,\tau_2,k)
=2\,k^2\,\Gamma^2\int_{-1}^{1}\!dz\!\int_{0}^{\infty}\!dp\;
\frac{p^2}{\tilde p^{4}}\,(1-z^2)^2\,
E_{\rm kin}(\tau_1,\tau_2,p)\,E_{\rm kin}(\tau_1,\tau_2,\tilde p)\,,
\end{equation}
which is the form used in the spectrum integral \cref{eq:Omega-production}.

\paragraph{Linear hydrodynamics for $(u,\lambda)$.}
Let $\lambda\equiv (\rho-\bar\rho)/\bar w$ be the normalized energy density perturbation. Linearized continuity and Euler equations give
\begin{equation}
\label{eq:ssm:lin-eqs}
\lambda'(\tau,\mathbf k) - i\,\mathbf k\!\cdot\!\mathbf u(\tau,\mathbf k)=0,
\qquad
\mathbf u'(\tau,\mathbf k) - i\,c_s^2\,\mathbf k\,\lambda(\tau,\mathbf k)=0\,,
\end{equation}
which imply harmonic evolution with dispersion $\omega=c_s k$. 
The velocity field $\mathbf{u}(\tau,\mathbf k)$ can be written in terms of mode amplitudes $A_\pm(k)$ fixed at the onset of the acoustic stage $\tau_*$ as
\begin{align}
\label{eq:ssm:Apm}
\mathbf{u}(\tau,\mathbf k)&= \hat{k}\,\sum_{s=\pm}A_s(k)\,e^{is\omega(\tau-\tau_*)}
\,,&
A_\pm(k)&=\tfrac12\Big[u(\tau_*,k)\pm c_s\,\lambda(\tau_*,k)\Big]\,.
\end{align}

\paragraph{Single-bubble kernels and amplitudes.}
The longitudinal velocity and energy contrast fields can be written in position space as a linear superposition of spherically symmetric single-bubble profiles \(v_{\rm ip}(\xi)\) and \(\lambda_{\rm ip}(\xi)\) nucleated at locations \(\mathbf{x}_n\) with lifetime \(T_n=\tau-\tau_n\) for the $n$-th bubble,
\begin{equation}
    \mathbf{u}(\tau,\mathbf{x}) = \sum_{n=1}^{N_b} \frac{\mathbf{x}-\mathbf{x}_n}{|\mathbf{x}-\mathbf{x}_n|} \,v_{\rm ip}(\xi_n)
    \qquad\text{and}\qquad
    \lambda(\tau,\mathbf{x}) = \sum_{n=1}^{N_b} \,\lambda_{\rm ip}(\xi_n) 
    \qquad\text{with}\qquad\xi_n \equiv\frac{|\mathbf{x}-\mathbf{x}_n|}{T_n}\,.
\end{equation}
In momentum space, this leads to a factorized form for the mode amplitudes~\cite{Hindmarsh:2016lnk,Hindmarsh:2019phv},
\begin{equation}
\label{eq:ssm:A-ensemble}
A_\pm(\mathbf k)=\sum_{n=1}^{N_b}\,\mathcal A_\pm(k T_n)\,T_n^3\,e^{i\mathbf k\cdot \mathbf x_n}
\qquad\text{with}\qquad
\mathcal A_\pm(\chi)=-\frac{i}{2}\Big[f'(\chi)\pm i\,c_s\,\ell(\chi)\Big]\,,
\end{equation}
where single-bubble functions $\mathcal A_\pm$ are conveniently expressed in terms of radial kernels built from the spherically symmetric profiles $v_{\rm ip}(\xi)$ and $\lambda_{\rm ip}(\xi)$,
\begin{equation}
\label{eq:ssm:f-l-kernels}
f(\chi)=\frac{4\pi}{\chi}\int_0^\infty d\xi\,v_{\rm ip}(\xi)\,\sin(\chi\xi),\qquad
\ell(\chi)=\frac{4\pi}{\chi}\int_0^\infty d\xi\,\xi\,\lambda_{\rm ip}(\xi)\,\sin(\chi\xi)\,.
\end{equation}
These kernels encode the two characteristic length scales of the SSM: the mean bubble separation $R_*$ and the shell thickness $\Delta_w\sim |\,\xi_w-c_s\,|/c_s$ set by the proximity of the wall speed to the sound speed.

\paragraph{Velocity UETC and its spectral decomposition.}
After averaging over nucleation centers and lifetimes, the ensemble average in \cref{eq:ssm:uu-UETC} can be calculated following Refs.~\cite{Hindmarsh:2016lnk,Hindmarsh:2019phv,RoperPol:2023dzg}, and the kinetic UETC takes the form
\begin{subequations}\begin{align}
\label{eq:ssm:Ekin-decomp}
E_{\rm kin}(\tau_1,\tau_2,k) &=
E_{\rm kin}^{(1)}(k)\,\cos\!\big[\omega\tau_-\big]+
E_{\rm kin}^{(2)}(k)\,\cos\!\big[\omega(\tau_+-2\tau_*)\big]+
E_{\rm kin}^{(3)}(k)\,\sin\!\big[\omega(\tau_+-2\tau_*)\big]\,,
\\
\label{eq:ssm:Ekin-final}
&\simeq E_{\rm kin}(k)\,\cos\!\big[\omega\tau_-\big]\,,
\end{align}\end{subequations}
where \(\tau_\pm\equiv\tau_1\pm\tau_2\) and we dropped the terms oscillating with $\tau_*$ in the second step, taking $E_{\rm kin}(k) = E_{\rm kin}^{(1)}(k)$.
The three components are explicit integrals over the bubble lifetime distribution and the single-bubble kernels. Writing $\tilde T\equiv \beta T$ and $K\equiv k/\beta$, where $\beta = (8 \pi)^3 \xi_w/R_*$ with the mean bubble separation $R_*$ is the nucleation rate parameter, and introducing the bubble lifetime distribution function $\nu(\tilde{T})$, one can express them as
\begin{equation}
\label{eq:ssm:Ekin-En}
E_{\rm kin}^{(n)}(k)=\frac{k^2}{2\pi^2\,\beta^6\,R_*^3}\int_0^\infty d\tilde T\; \nu(\tilde T)\;{\mathcal E}^{(n)}(K\tilde T)\,.
\end{equation}
Two nucleation scenarios are useful benchmarks
\begin{equation}
\label{eq:ssm:nucleation-weights}
\text{exponential:}\quad  \nu_{\rm exp}(\tilde T)=e^{-\tilde T},\qquad
\text{simultaneous:}\quad  \nu_{\rm sim}(\tilde T)=\tfrac12\,\tilde T^2\,e^{-\tilde T^3/2}.
\end{equation}

The building blocks in \cref{eq:ssm:Ekin-En} ${\mathcal E}^{(n)}$ are
\begin{align}
\label{eq:ssm:calE}
{\mathcal E}^{(1)}(\chi)=\big|\mathcal A_+(\chi)\big|^2 \,,\qquad 
{\mathcal E}^{(2)}(\chi)=\Re\!\big[\mathcal A_+(\chi)\mathcal A_-^*(\chi)\big]\,,\qquad 
{\mathcal E}^{(3)}(\chi)=\Im\!\big[\mathcal A_+(\chi)\mathcal A_-^*(\chi)\big]\,.
\end{align}

Causality of a longitudinal (irrotational) flow enforces $E_{\rm kin}\propto k^4$ as $k\to 0$, while the shell thickness yields a $k^{-2}$ decay at large $k$. For $k\gg(2\,\delta\tau_{\rm fin})^{-1}$, only the stationary term in \cref{eq:ssm:Ekin-decomp} matters inside the time integrals and we define the \emph{kinetic spectrum} $E_{\rm kin}(k)\equiv E_{\rm kin}^{(1)}(k)$. However, in the strict $k\to 0$ limit the combination $E_{\rm kin}^{(1)}+E_{\rm kin}^{(2)}$ must be retained to obtain the correct causal infrared behavior of the GW spectrum~\cite{RoperPol:2023dzg}.

\begin{figure}[t]
    \centering
    \includegraphics{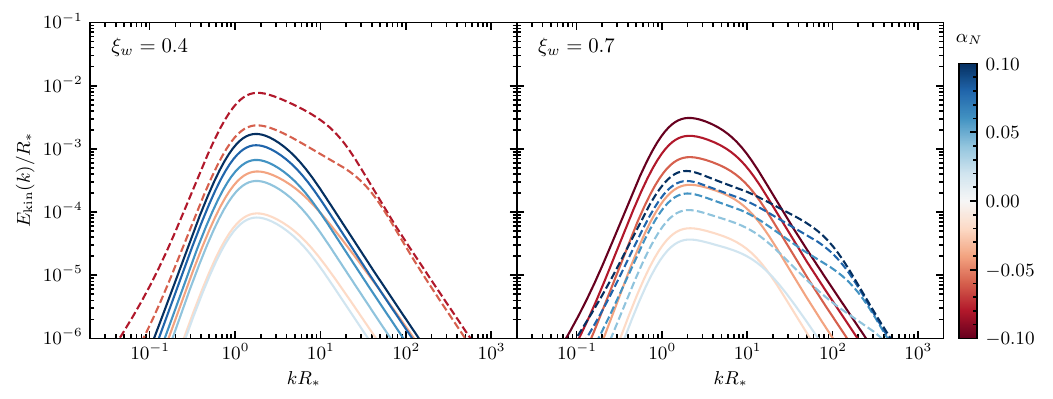}
    \caption{\textbf{Kinetic spectrum.} Time-independent component $E_{\rm kin}(k)/R_*$ of the fluid velocity UETC for wall speeds of $\xi_w=0.4$ (left) and $\xi_w=0.7$ (right), comparing direct (blue) and inverse (red) branches. Solid lines correspond to (direct or inverse) deflagrations and detonations, whereas dashed lines are hybrids. The corresponding values of $\alpha_N$ are indicated by the respective colors, where shades of blue (red) are direct (inverse) transitions.}
    \label{fig:Ekin}
\end{figure}

\begin{figure}[t]
    \centering
    \includegraphics{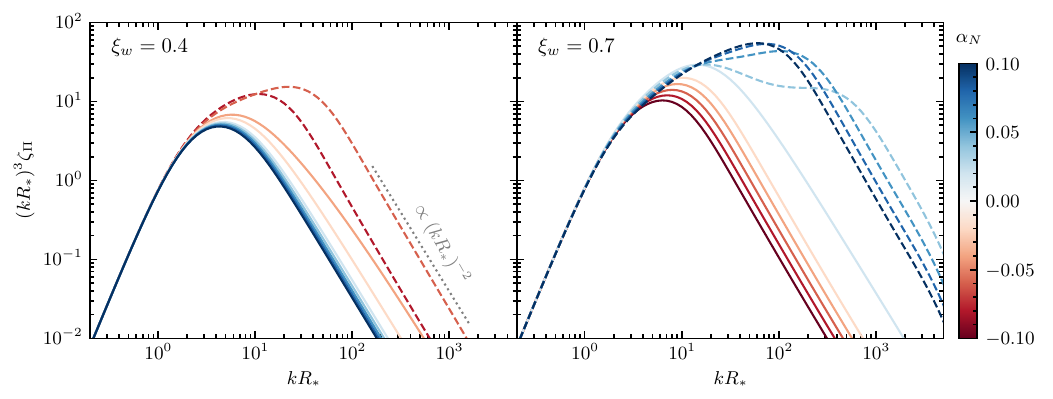}
    \caption{\textbf{Anisotropic-stress UETC spectral shape.} The combination $K^3\zeta_\Pi(K)$, \cref{eq:ssm:Epi-final,eq:ssm:zetapi}, controls the position and height of the GW peak. The ultraviolet behavior corresponds to $\zeta_\Pi\propto K^{-5}$. Solid lines correspond to deflagration or detonation type solutions, whereas dashed lines are hybrids. The coloring encodes the corresponding values of $\alpha_N$, where shades of blue and red indicate direct and inverse dynamics, respectively.}
    \label{fig:zeta_Pi}
\end{figure}

\begin{figure}
    \begin{minipage}[t]{.5\linewidth}
    \centering
    \includegraphics{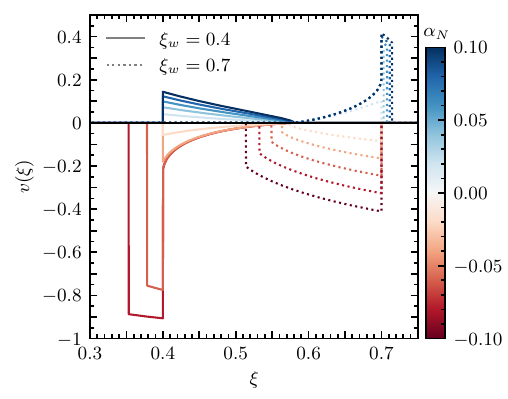}
    \parbox{.95\linewidth}{%
        \caption{{\bfseries Fluid profiles} corresponding to the ones used to produce \cref{fig:gw_xiw_04_07}. Solid and dotted lines correspond to a wall velocity of $\xi_w=0.4$ and $\xi_w=0.7$, respectively, and the colors indicate the value of the strength parameter $\alpha_N$.}
        \label{fig:velocity_profiles}
    }
    \end{minipage}%
    \begin{minipage}[t]{.5\linewidth}
    \centering
    \includegraphics{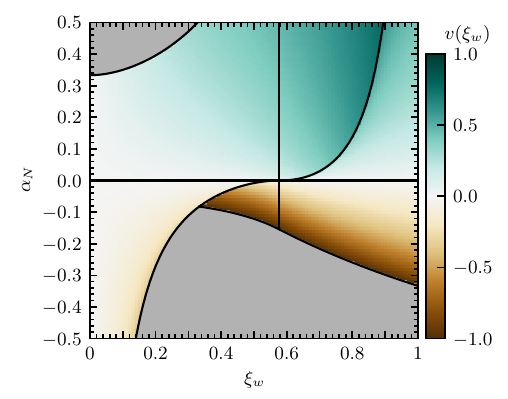}
    \parbox{.95\linewidth}{%
        \caption{{\bfseries Maximal fluid velocities} at the wall position, $v(\xi_w)$, in the $(\xi_w, \alpha_N)$ plane. The sound-shell model assumes non-relativistic fluid velocities.}
        \label{fig:peak_velocities}
    }
    \end{minipage}
\end{figure}

\paragraph{From $E_{\rm kin}$ to the anisotropic-stress power spectrum.}
Using \cref{eq:ssm:Ekin-final,eq:ssm:Epi-final} we can obtain the anisotropic stress power spectrum $k E_\Pi(k) \equiv k \,E_\Pi(\tau,\tau,k)$ at equal times.
Introducing the dimensionless variables $K\equiv kR_*$, $P\equiv pR_*$ and $\tilde P\equiv \tilde p R_*=\sqrt{K^2+P^2-2KPz}$, and factorizing the equal-time kinetic spectrum%
\footnote{%
    Note the different normalization of the dimensionless kinetic spectrum $\zeta_{\rm kin}$ compared to Ref.~\cite{RoperPol:2023dzg}, where it is normalized to unit peak amplitude.
}
as $E_{\rm kin}(k)=R_* \bar{U}_f^2\,\zeta_{\rm kin}(K)$, with peak amplitude $E_{\rm kin}^*$ and shape $\zeta_{\rm kin}(K)$, we obtain
\begin{equation}
\label{eq:ssm:Epi-factor}
k\,E_\Pi(k)\ \simeq\
2\, K^3\,
\Omega_K^{2}\,\mathcal C\,\zeta_\Pi(K)\,,
\end{equation}
with a numerical normalization $\mathcal C$ chosen so that $\zeta_\Pi\to 1$ as $K\to 0$. The \emph{dimensionless} shape $\zeta_\Pi(K)$ is obtained by the angular-momentum convolution
\begin{equation}
\label{eq:ssm:zetapi}
\mathcal C\,\zeta_\Pi(K)=
\int_0^\infty dP\,P^2\,\zeta_{\rm kin}(P)
\int_{-1}^{1}dz\,(1-z^2)^2\,\frac{\zeta_{\rm kin}(\tilde P)}{\tilde P^{4}},
\qquad
\mathcal C\equiv \frac{16}{15}\int_0^\infty dK\,\frac{\zeta_{\rm kin}^2(K)}{K^2}\,.
\end{equation}
\Cref{eq:ssm:zetapi} makes manifest three robust properties: $\zeta_\Pi(K)$ is monotonically decreasing from unity for increasing values of $K$; the combination $K^3\zeta_\Pi(K)$ peaks near $K\simeq K_{\rm GW}=\mathcal O(1)$ set by the sound-shell thickness and $R_*$; and the ultraviolet tail is universal, $\zeta_\Pi(K)\propto K^{-5}$, i.e.\ four powers steeper than the large-$K$ decay of $\zeta_{\rm kin}$. \Cref{fig:Ekin,fig:zeta_Pi} show different profiles, presented in \cref{fig:velocity_profiles}, for $\zeta_{\rm kin}$ and $\zeta_{\Pi}$, respectively, in the range $\alpha_N \in [-0.1, 0.1]$ for two choices of the wall velocity $\xi_w=0.4, 0.7$.

\subsection{Time kernel and gravitational wave spectrum}
\label{sec:time-kernel}

Cosmic expansion enters the acoustic GW calculation through the double time integral of the source Green’s functions weighted by the $1/(\tau_1\tau_2)$ kernel. Inserting \cref{eq:ssm:Epi-final,eq:ssm:Ekin-final} into \cref{eq:Omega-production} and factoring out the stationary angular structure of the anisotropic stress and collecting the time dependence into a single object, we obtain
\begin{equation}
    \label{eq:OmegaGW-full-prod}
    \Omega_{\rm GW}^*(K) =
    3\,K^3\,\Omega_K^2\,
    \int\limits_{0}^{\infty}\!\dd P \int\limits_{-1}^{1}\!\dd z\,
        \frac{P^2}{\tilde P^{4}}\,(1-z^2)^2\,
        \zeta_{\rm kin}(P)\,\zeta_{\rm kin}(\tilde P)\,
    \Delta\!\left(\frac{\delta\tau_{\rm fin}}{R_*};K,P,\tilde P\right),
\end{equation}
where we defined the time kernel function
\begin{equation}
\label{eq:Delta-def}
\Delta(\delta\tau_{\rm fin};k,p,\tilde p)\;\equiv\;
\int_{\tau_*}^{\tau_{\rm fin}}\!\frac{d\tau_1}{\tau_1}
\int_{\tau_*}^{\tau_{\rm fin}}\!\frac{d\tau_2}{\tau_2}\;
\cos(c_s p\,\tau_-)\,\cos(c_s \tilde p\,\tau_-)\,\cos(k\,\tau_-)\,,
\qquad \tau_- \equiv \tau_2-\tau_1,
\end{equation}
with $\delta\tau_{\rm fin}\equiv\tau_{\rm fin}-\tau_*$. 

\paragraph{Time kernel evaluation.}
Using product-to-sum identities and separating the $\tau_1$ and $\tau_2$ dependences, the integral can be written as a sum over four combined frequencies,
\begin{equation}
\label{eq:Delta-sum}
\Delta(\delta\tau_{\rm fin};k,p,\tilde p)\;=\;\sum_{m,n=\pm 1}\,\Delta_{mn}\!\left(\delta\tau_{\rm fin};\hat p_{mn}\right),
\qquad
\hat p_{mn}\;\equiv\; c_s\,(p+m\,\tilde p)+n\,k\,,
\end{equation}
where each term admits a closed form in terms of cosine and sine integral functions. Introducing the shorthand
\begin{equation}
\label{eq:DeltaCiDeltaSi-def}
\Delta{\rm Ci}(\tau,\hat p)\;\equiv\;{\rm Ci}(\hat p\,\tau)-{\rm Ci}(\hat p\,\tau_*)\,,
\qquad
\Delta{\rm Si}(\tau,\hat p)\;\equiv\;{\rm Si}(\hat p\,\tau)-{\rm Si}(\hat p\,\tau_*)\,,
\end{equation}
the exact result for each $(m,n)$ contribution is
\begin{equation}
\label{eq:Delta-mn}
\Delta_{mn}\!\left(\delta\tau_{\rm fin};\hat p_{mn}\right)
\;=\;\frac{1}{4}\Big[
\Delta{\rm Ci}(\tau_{\rm fin},\hat p_{mn})^2
+\Delta{\rm Si}(\tau_{\rm fin},\hat p_{mn})^2
\Big].
\end{equation}
\Cref{eq:Delta-sum,eq:DeltaCiDeltaSi-def,eq:Delta-mn} implement the expansion history exactly during radiation domination (where $a''/a=0$ in conformal time) and make manifest two parametric regimes controlled by the magnitude of the combined frequencies $\hat p_{mn}$ relative to $\mathcal{H}_*\equiv 1/\tau_*$: for $\hat p_{mn}\ll \mathcal{H}_*$ one has $\Delta{\rm Ci}\!\to\!\ln(\tau_{\rm fin}\mathcal{H}_*)$ and $\Delta{\rm Si}\!\to\!0$, yielding $\Delta\sim \ln^2(\tau_{\rm fin}\mathcal{H}_*)=\ln^2(1+\mathcal{H}_*\,\delta\tau_{\rm fin})$; for a short-lived source in an almost-flat spacetime, $\mathcal{H}_*\,\delta\tau_{\rm fin}\ll 1$, one recovers the familiar quadratic growth $\Delta\sim (\mathcal{H}_*\,\delta\tau_{\rm fin})^2$. 

\paragraph{Infinite duration limit.}
In the limit of an infinite duration of the source, $k\,\delta\tau_{\rm fin}\to\infty$, the contributions $\Delta_{mn}$ in \cref{eq:Delta-mn} to the time kernel can be simplified to~\cite{RoperPol:2023dzg}
\begin{align}
    \label{eq:DeltamnInfiniteDuration}
    \Delta_{mn}\!\left(\delta\tau_{\rm fin};\hat p_{mn}\right) \;\xrightarrow{k\,\delta\tau_{\rm fin}\to\infty}\;
    \frac{\pi}{2} \Upsilon(\delta\tau_{\rm fin})\,\delta(\hat{p}_{mn}/\mathcal{H}_*)
    \qquad\text{with}\qquad
    \Upsilon(\delta\tau_{\rm fin}) = \frac{\mathcal{H}_*\delta\tau_{\rm fin}}{1 + \mathcal{H}_*\delta\tau_{\rm fin}}\,.
\end{align}
Plugging \cref{eq:DeltamnInfiniteDuration} into \cref{eq:OmegaGW-full-prod}, we obtain the simplified spectrum derived in Ref.~\cite{Hindmarsh:2019phv},
\begin{equation}
    \label{eq:OmegaGW-long}
    \Omega_\mathrm{GW}^*(K) =
    \frac{3 \pi}{2 c_s}\,\mathcal{H}_* R_*\,\Upsilon(\delta\tau_{\rm fin})\,K^2\,\Omega_K^2\,
    \int\limits_{P_-}^{P_+}\!\dd P\,
        \frac{P}{\tilde P^{3}}\,(1-z^2)^2\,
        \zeta_{\rm kin}(P)\,\zeta_{\rm kin}(\tilde P)\,,
\end{equation}
with $P_\pm = \frac{1}{2} K (1\pm c_s)/c_s$ and $\tilde P = K/c_s - P$.

\paragraph{Present-day gravitational wave spectrum.}
The gravitational wave spectrum today is related to the spectrum at production via
\begin{equation}
    \label{eq:OmegaGW-full}
    \Omega_{\rm GW}(f) = T_{\rm GW}\, \Omega_{\rm GW}^*(K)
    \qquad\text{with}\qquad
    f = \frac{H_{*,0}}{2\pi}\,\frac{K}{\mathcal{H}_* R_*} \, ,
\end{equation}
where $\Omega_\mathrm{GW}^*$ is given in \cref{eq:OmegaGW-full-prod} (or \cref{eq:OmegaGW-long} in the infinite duration limit), and the Hubble rate at production red-shifted to today is
\begin{equation}
    H_{*,0} = \frac{\mathcal{H}_*}{a_0} 
    = \SI{1.65e-5}{\Hz} \,\left(\frac{g_*}{100}\right)^\frac{1}{6}\,\left(\frac{T_*}{\SI{100}{\GeV}}\right)\,.
\end{equation}
The gravitational-wave spectrum is hence determined in terms of the strength parameter~$\alpha_N$, the wall velocity $\xi_w$, the average bubble size $\mathcal{H}_* R_*$ normalized to the Hubble rate, the transition temperature $T_*$ and duration $\mathcal{H}_* \delta\tau_\mathrm{fin}$. The latter is commonly estimated from the eddy-turnover time at the injection scale, using the root-mean-square velocity defined in \cref{eq:ssm:OmegaK_and_Urms},
\begin{equation}
    \mathcal{H}_* \delta\tau_{\rm fin} \sim \frac{\mathcal{H}_* R_*}{\bar{U}_f}\,.
\end{equation}

\section{Results: Distinguishing direct and inverse gravitational-wave signals}
\label{sec:results}

\begin{figure}
    \centering
    \includegraphics{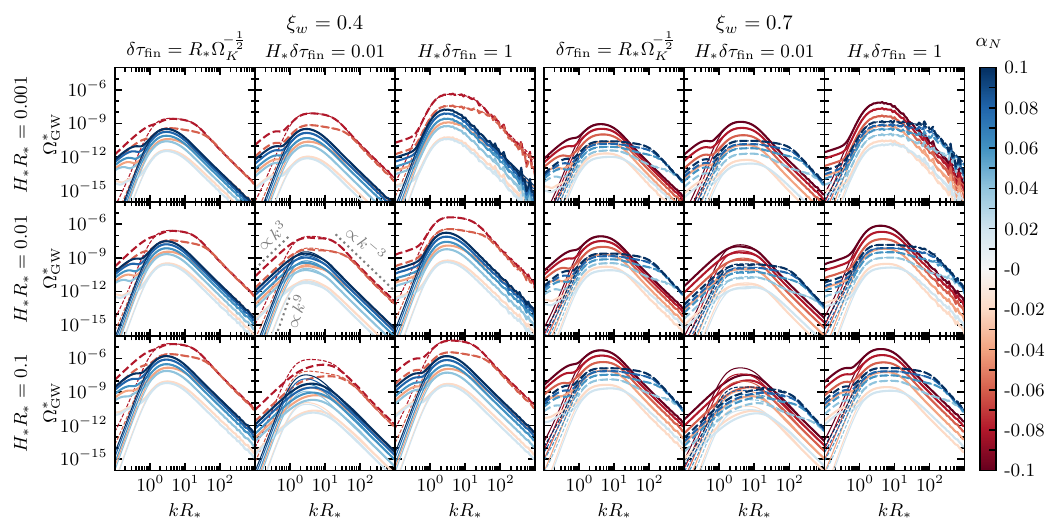}
    \caption{{\bfseries Gravitational-wave spectra at production} for $\xi_w=0.4$ (left) and $\xi_w=0.7$ (right) and for different values of $\alpha_N$, varying the acoustic lifetime $\delta\tau_{\rm fin}$ and the mean separation $R_*$. Blue colors refer to direct PT, while red ones to inverse PT. The causal $k^3$ rise at very small $k$, a possible linear intermediate regime, and the $k^{-3}$-controlled high-$k$ decay are visible across the panels. A steeper near-peak shoulder appears for long-lived sources in Hubble units. Solid lines show deflagration/detonation while dashed lines show hybrids. Thin lines refer to the infinite source duration as in the original SSM.
    }
    \label{fig:gw_xiw_04_07}
\end{figure}

The inputs assembled above determine the present-day gravitational-wave spectrum through \cref{eq:OmegaGW-full}. At fixed $(\xi_w,\alpha_N)$ the single-bubble profiles set the shell geometry and thus the shapes $\zeta_{\rm kin}(K)$ and $\zeta_\Pi(K)$; $R_*$ fixes the frequency scale, $\Omega_K$ the normalization, and $H_*\delta\tau_{\rm fin}$ the degree of temporal coherence. \Cref{fig:Ekin,fig:zeta_Pi} provide the kinetic spectrum $E_{\rm kin}$ and the anisotropic-stress power-spectral shape $K^3 \zeta_\Pi(K)$ of the model for representative wall speeds, while \cref{fig:gw_xiw_04_07} displays the resulting gravitational-wave spectra at production for different lifetimes and mean separations. 
As the fluid profiles enter the gravitational-wave spectrum in \cref{eq:OmegaGW-full-prod} only through the kinetic spectrum $E_\mathrm{kin}$, \cref{fig:Ekin,fig:zeta_Pi} already provides an indicator for whether signals from inverse and direct transitions can be distinguished. While the shape of the kinetic spectrum changes with the input parameters, similar shapes can be realized for both direct and inverse transitions. This implies that discrimination based on the GW signal will be challenging.

\begin{figure}
    \centering
    \includegraphics{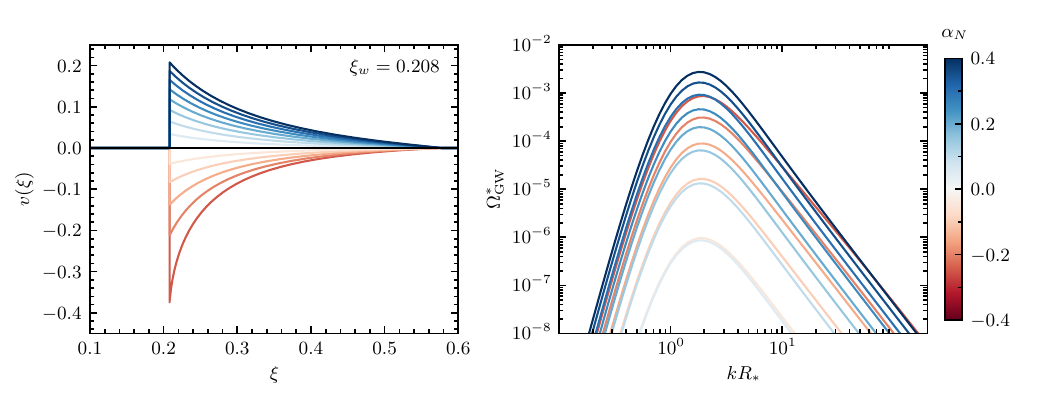}
    \caption{{\bfseries Fluid profiles and gravitational-wave spectrum at production} for $\xi_w=0.208$, scanning the strength $\alpha_N$ across the range where direct and inverse solutions exist. The peak tracks $K_{\rm GW}\sim\mathcal{O}(1)$ and the overall normalization follows $\Omega_K^2$ and $H_*R_*$.}
    \label{fig:xi0208}
\end{figure}

Furthermore, at first glance, one might conclude from \cref{fig:gw_xiw_04_07} that inverse transitions systematically yield a louder signal at fixed \((\mathcal H_*R_*,\mathcal H_*\delta\tau_{\rm fin})\). This impression, however, reflects the \emph{chosen slice} of parameter space: in the range \(\alpha_N\in[-0.1,0.1]\) the inverse branch sits close to its hydrodynamic boundary at the most negative allowed \(\alpha_N\). Proximity to this bound boosts the maximal fluid velocity achieved at the bubble wall, $v(\xi_w)$, as can be appreciated from \cref{fig:velocity_profiles}, and therefore the kinetic energy fraction \(\Omega_K\), raising the overall amplitude. When the slice is widened to scan the entire viable range of $\alpha_N$ at fixed wall velocity, e.g\ \(\alpha_N\in[-0.25,0.4]\) at \(\xi_w=0.208\) as shown in \cref{fig:xi0208}, the direct branch also reaches its maximal admissible strength and, correspondingly, attains a comparable --- and in places larger --- signal. In short, the apparent hierarchy between inverse and direct signals is not universal: it tracks how close each branch lies to its \((\alpha_N,\xi_w)\) boundary, which controls \(\kappa\), \(\Omega_K\), and thus the peak normalization at fixed \((\mathcal H_*R_*,\mathcal H_*\delta\tau_{\rm fin})\).

As a word of caution, it should be noted that the absolute value of the lower bound on $\alpha_N$ is smaller than the corresponding upper bound (cf.\ \cref{fig:hydro_solution_space}) for most values of the wall velocity, as the value $\alpha_+$ entering the matching conditions in \cref{eq:junctionAB} at the wall is always smaller (i.e.\ more negative for negative values) than the value $\alpha_N$.
As a consequence, inverse transitions tend to have larger fluid velocities compared to direct transitions at the same absolute value of $\alpha_N$.
Furthermore, in inverse deflagrations and hybrids, relativistic fluid velocities at the position of the wall can be reached even at moderate values of $\alpha_N \simeq -\mathcal{O}(0.1)$, as shown in \cref{fig:peak_velocities}.
The sound shell model hence breaks down in this regime, as \cref{eq:ssm:TT-def} assumes the non-relativistic limit.
The same applies for direct hybrids close to the Jouguet boundary.

Nonetheless, several robust, physically transparent contrasts emerge. In the infrared, both direct and inverse transitions exhibit the same causal $k^3$ rise once expansion is included;  there are only differences in the overall amplitude of the signal, controlled by $\Omega_K^2$ and by the slow $\ln^2(1+\mathcal H_*\delta\tau_{\rm fin})$ growth encoded in the time kernel. Around the maximum, the shell thickness $\Delta_w\sim|\xi_w-c_s|/c_s$ becomes decisive: as $\xi_w\to c_s$ the shell narrows and an intermediate, almost linear shoulder can appear between the causal tail and the peak. 
Near the peak, long-lived sources ($\mathcal H_*\delta\tau_{\rm fin}\gg 1$) develop a sharp intermediate-frequency rise that can locally exhibit $k^9$ growth; expansion smooths this feature compared to a strictly stationary source, but its prominence remains branch dependent through the single-bubble kernels. On the ultraviolet side, both branches share the same $k^{-3}$ decay fixed by the anisotropic-stress convolution and therefore offer no discriminating power there.

To quantify these qualitative observations and assess more systematically to what extent the gravitational-wave spectra from direct and inverse transitions can be distinguished, we now turn to a shape-based comparison using spectral angles.

\subsection{Spectral angles: from pure shape to amplitude-aware comparisons}

\noindent To assess in how far the gravitational-wave signals can be disentangled across direct and inverse phase transitions, we now turn to building two different measures.

The present-day peak frequency is primarily set by the mean bubble size $\mathcal{H}_* R_*$ and through red-shifting by the transition temperature $T_*$. The amplitude, on the other hand, depends on $\alpha_N$ and $\xi_w$ via the fluid profiles and on $\delta\tau_{\rm fin}/R_*$ from the time kernel \cref{eq:Delta-def}, but is to independent of the temperature (neglecting changes in the effective number of degrees of freedom).
The spectral shape, in turn, primarily depends on $\xi_w$ and $\alpha_N$, up to $\mathcal{H}_*\delta\tau_{\rm fin}$ and $\mathcal{H}_*R_*$ dependent modification in the low-frequency tail.
For a given fluid profile determined by $(\xi_w,\alpha_N)$, it is thus always possible to shift the peak frequency to arbitrary values using the temperature, and, within limits, to adjust the peak amplitude using $\mathcal{H}_* R_*$ and $\mathcal{H}_* \delta\tau_{\rm fin}$ without major effects on the spectral shape.
We can therefore expect the spectral shape of the gravitational wave signal to be the primary handle to disentangle the input parameters $\xi_w$ and $\alpha_N$, and hence to distinguish direct and inverse PT signals.

In the following, we therefore use the spectral angle~\cite{Kruse:1993145} to evaluate how far two spectra can be discriminated based on their shape.
As the wall velocity $\xi_w$ and strength parameter $\alpha_N$ also significantly influence total energy in GWs through $\Omega_K$, we further introduce a \emph{modified} spectral angle, to indicate in how far differences in the amplitude can help the spectral discrimination at fixed $\mathcal{H}_* R_*$ and $\mathcal{H}_*\delta\tau_{\rm fin}$.
In both cases, we take all spectra to be peaked at the same frequency (i.e.\ adjusting the transition temperature accordingly), and hence use the shifted spectra $\bar\Omega(x)$ in terms of the dimensionless frequency $x$ normalized to the peak position in the following,
\begin{equation}
    \bar{\Omega}(x) \;\equiv\; \Omega_\mathrm{GW}(x f_\mathrm{peak})
    \qquad\text{with}\qquad
    x \;\equiv\; {f}/{f_\mathrm{peak}}\,,
\end{equation}
so that the spectra peak at $x=1$.

\paragraph{Standard spectral angle.}
The spectral angle assesses how similar or different two spectra $\bar{\Omega}_1$ and $\bar{\Omega}_2$ are by treating them as vectors in frequency space and evaluating the angle between the two vectors. A spectral angle of $\theta_{12} = 0$ indicates that the spectra have identical shape, where as an angle of $\theta_{12} = \pi/2$ corresponds to spectra with non-overlapping support in frequency space ($\cos\theta_{12} < 0$ would require negative spectra).

The \emph{standard} spectral angle \(\theta_{12}\) is then
\begin{equation}
    \label{eq:theta-standard-L2}
    \cos\theta_{12} \;=\; \frac{\big(\bar{\Omega}_1,\bar{\Omega}_2\big)}{\big\|\bar{\Omega}_1\big\|_2\,\big\|\bar{\Omega}_2\big\|_2\rule{0pt}{11pt}}\,,
    \qquad\text{with}\qquad
    \big(\bar{\Omega}_i,\bar{\Omega}_j\big) \!\equiv\! \int\!\dd\log x\;\bar{\Omega}_i(x)\,\bar{\Omega}_j(x)
    \quad\text{and}\quad
    \big\|\bar{\Omega}_i\big\|_2 \;\equiv\; \sqrt{\big(\bar{\Omega}_i,\bar{\Omega}_i\big)}\,,
\end{equation}
where $\big(\bar{\Omega}_i,\bar{\Omega}_j\big)$ denotes the scalar product and $\big\|\bar{\Omega}_i\big\|_2$ the corresponding \(L^2\) norm over \(\dd\log x\).
This construction is symmetric under $(1\leftrightarrow2)$ and insensitive to overall amplitude rescalings; \(\theta_{12}\) probes only widths, asymmetries, and slopes of the spectra.

\begin{figure}[t]
  \centering
  \includegraphics[width=\linewidth]{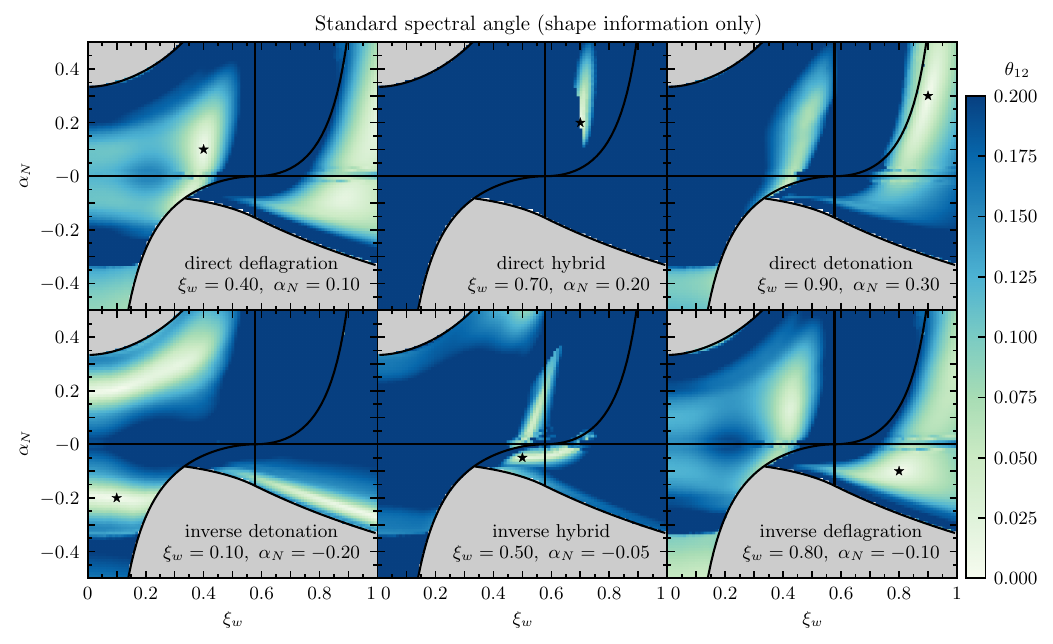}
  \caption{\textbf{Standard spectral angle} \(\theta_{12}\) from \cref{eq:theta-standard-L2}.
All spectra are self-aligned in frequency so that their peaks sit at \(x\!=\!1\), and each curve is
\(L^2(d\!\log x)\)-normalized. Each panel fixes one \((\xi_w,\alpha_N)\) as the
reference spectrum (black star) and scans the rest of the plane. The color scale
encodes \(\theta_{12}\in[0,\pi/2]\): smaller angles indicate shapes that closely
match the reference. Because peak
location and overall amplitude are removed, many regions appear nearly indistinguishable --- direct vs.\
inverse PTs, and even nearby direct solutions, often yield small \(\theta_{12}\), illustrating the
strong shape degeneracy of the standard metric.}
  \label{fig:sam:standard}
\end{figure}
\begin{figure}
    \centering
    \includegraphics[width=\linewidth]{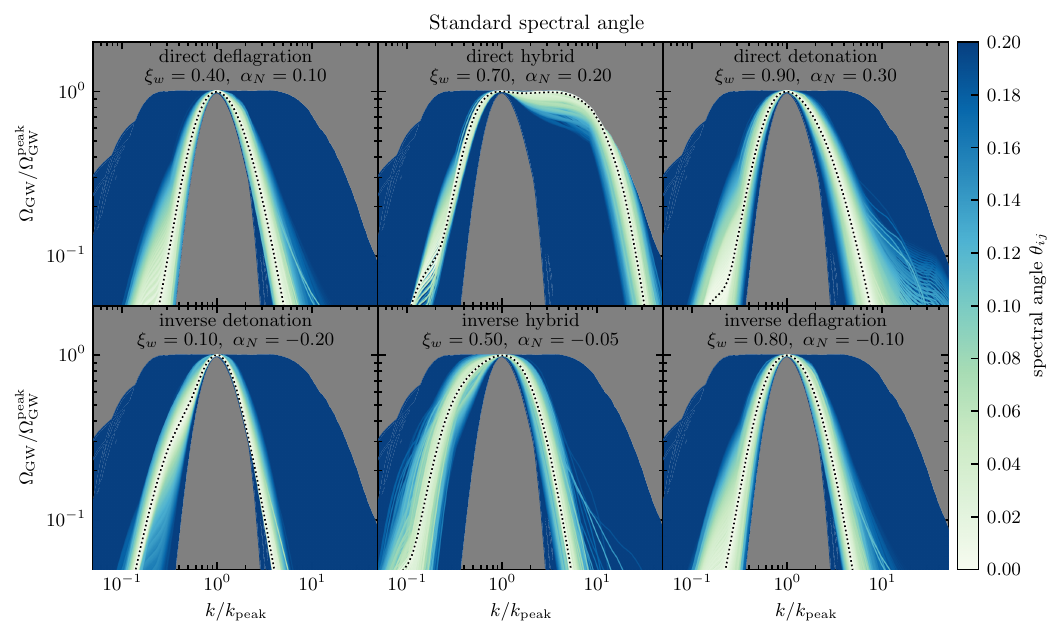}
    \caption{{\bfseries \boldmath Gravitational-wave spectra $\Omega_\mathrm{GW}^*$ and (standard) spectral angle $\theta_{12}$.} The color encodes the corresponding spectral angles with respect to the indicated reference spectra (black dotted line). The spectra are normalized to unit amplitude and correspond to the range of $\xi_w$ and $\alpha_N$ explored in \cref{fig:sam:standard}.}
    \label{fig:GW_spectra}
\end{figure}

\Cref{fig:sam:standard} shows \(\theta_{12}\), comparing a fixed reference (black star) to all other points in the \((\xi_w,\alpha_N)\) plane. While the procedure isolates \emph{shape} information, the resulting maps remain highly degenerate: large regions exhibit small angles, making it difficult to distinguish direct from inverse transitions and, in many cases, even to discriminate among different direct (or different inverse) points. Only in limited corners of parameter space do the shapes differ enough to yield larger angles, underscoring that shape-only information provides limited discriminating power and motivates the amplitude-informed variants discussed later. 
The scan was performed by fixing \( \mathcal{H}_* \delta \tau_{\mathrm{fin}} = 1 \) and \( \mathcal{H}_* R_* = \sqrt{\Omega_K} \mathcal{H}_* \delta \tau_{\mathrm{fin}}\), ensuring that nonlinear effects set in approximately one Hubble time after the bubble collisions. We verified that the conclusion does not change qualitatively when using the approximate form \cref{eq:OmegaGW-long}, whose spectral shape depends on $\xi_w$ and $\alpha_N$ alone.

It should be emphasized that the spectral angle as defined in \cref{eq:theta-standard-L2} discriminates spectra predominantly based on the region around the peak; differences in the low- and high-frequency regions contribute less due to the suppressed amplitude in the tails.
This in particular means that, since we shifted the spectra to peak at the same frequency, we generally obtain low spectral angles.
A proper investigation of whether different spectra can be distinguished for a given  value of the spectral angle depends on assumptions on the experimental sensitivity and is left for future work.
To provide some intuition of the difference in spectral shape that corresponds to a value of the spectral angle, \cref{fig:sam:standard} shows the respective spectra (normalized to unit amplitude) with the corresponding value of $\theta_{12}$ indicated by the color coding.

\paragraph{Modified spectral angle.}
To lift part of this degeneracy while keeping the same peak alignment \(x=1\), we retain information about the \emph{relative overall yield}. 
We define the \emph{modified} spectral angle as
\begin{align}
  \cos\widetilde\theta_{12}\;=\;
  \frac{\big(\bar{\Omega}_1,\bar{\Omega}_2\big)}{\max\left(\big\|\bar{\Omega}_1\big\|_2,\,\big\|\bar{\Omega}_2\big\|_2\right)^{\!2}}
  \;=\; \frac{\bar{\Omega}_{\min}}{\bar{\Omega}_{\max}} \,\cos\theta_{12}\,,
  \qquad
  \bar{\Omega}_{\substack{\min\\\max}} \;\equiv\; \begin{cases}
      \min\left(\big\|\bar{\Omega}_1\big\|_2,\,\big\|\bar{\Omega}_2\big\|_2\right)\\
      \max\left(\big\|\bar{\Omega}_1\big\|_2,\,\big\|\bar{\Omega}_2\big\|_2\right)\rule{0pt}{12pt}
  \end{cases}\,,
  \label{eq:theta-modified}
\end{align}
which corresponds to the \emph{modified} denominator used in our scans.

\begin{figure}[t]
  \centering
  \includegraphics[width=\linewidth]{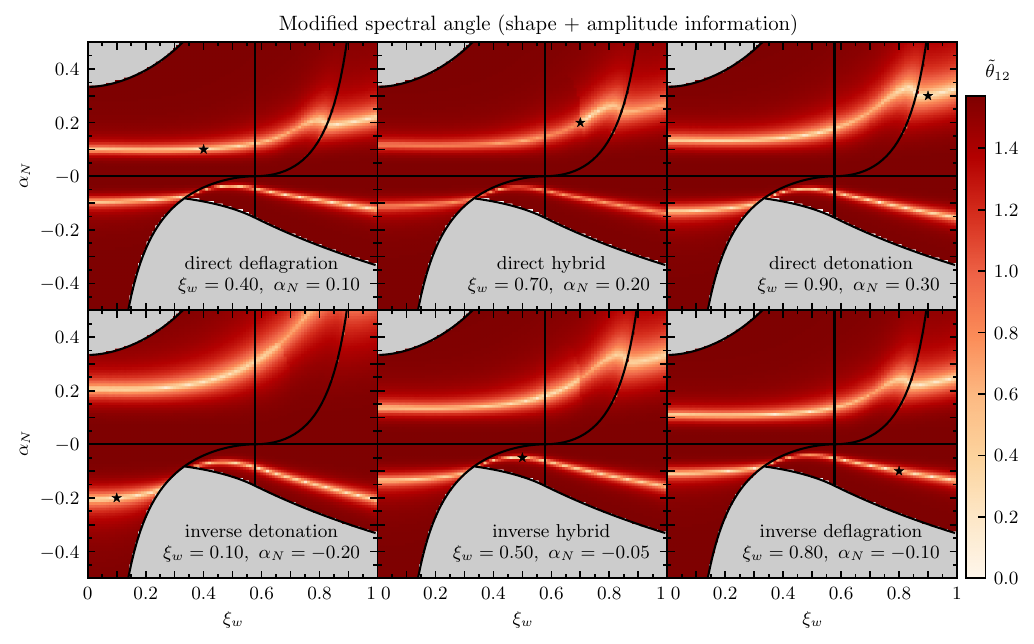}
  \caption{\textbf{Modified spectral angle} \(\widetilde\theta_{12}\) from \cref{eq:theta-modified} with \(A_i=\Omega_{K,i}\) (multi-bubble efficiency). Including amplitude information helps reduce degeneracies; the low-\(\xi_w\) band appears comparatively flat because \(\Omega_K\) varies slowly there.}
  \label{fig:sam:modified}
\end{figure}

\begin{figure}
    \centering
    \includegraphics[width=\linewidth]{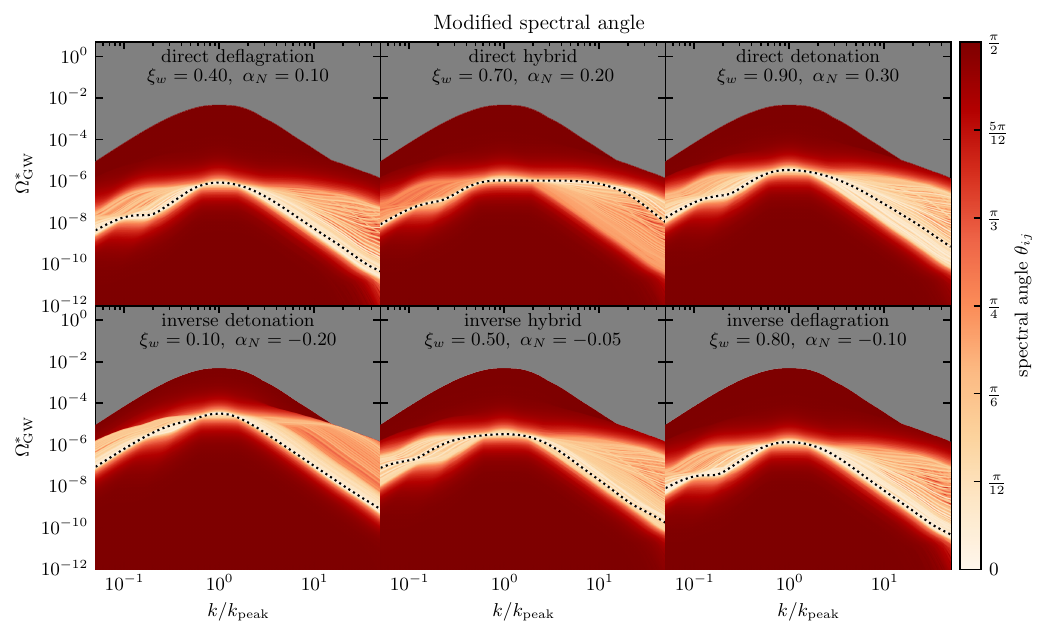}
    \caption{\textbf{Gravitational-wave spectra and modified spectral angle $\bar\theta_{12}$} The color encodes the corresponding modified spectral angles with respect to the indicated reference spectra (black dotted line). The spectra correspond to the range of $\xi_w$ and $\alpha_N$ explored in \cref{fig:sam:modified}.}
    \label{fig:GW_spectra_modified}
\end{figure}

The regions of parameter space giving spectra similar in both shape and amplitude are siginificantly different from the ones discriminated in shape only. 
\Cref{fig:sam:modified} depicts the modified spectral angle $\tilde{\theta}_{12}$ in the \((\xi_w,\alpha_N)\) plane (fixing $\mathcal{H}_*\delta\tau_{\rm fin}$ and $\mathcal{H}_* R_*$ as  before).
Still, one obtains degeneracies along filaments in the parameter plane spanning through different types of hydrodynamic solutions, both inverse and direct, again indicating that direct and inverse phase transitions exhibit similar spectra. 
Once more, it should however be emphasized that the question of which value of the (modified) spectral angle corresponds to spectra visibly different in data depends on the experiment under consideration and cannot be answered in general. Our qualitative conclusion, however, should hold.
For reference, the corresponding spectra are shown alongside the respective modified spectral angle are shown in \cref{fig:GW_spectra_modified}.

We highlight a distinctive low-velocity effect: the \emph{modified spectral angle} tends to flatten as the wall speed decreases. Tracing this back to the energy budget, we find that the multi-bubble kinetic fraction $\Omega_K$ also flattens in the $\xi_w\!\to\!0$ limit. A steeper suppression might have been expected by considering the single-bubble efficiency only, which is known to drop rapidly at small $\xi_w$ \cite{Barni:2024lkj}. As we argue in \cref{app:rms}, this is due to the fact that the overlap of many bubbles at collision compensates for the poor conversion of the isolated expanding bubbles.

\section{Conclusions and outlook}
\label{sec:conclusion}
Inverse phase transitions in the early Universe feature distinct fluid profiles compared to their direct counter-parts, potentially leaving an imprint in the corresponding stochastic gravitational wave background.
While direct transitions have been investigated abundantly in the literature, we here conduct a first study of the local-thermal-equilibrium properties --- giving an estimate of the value of the bubble wall velocity and mapping out the parameter space consistent with the LTE assumption --- and the gravitational-wave signal of first-order phase transitions with inverse hydrodynamics. 

Hydrodynamics alone does not fix the wall speed; to obtain a predictive estimate during the expansion we invoked local thermal equilibrium (LTE) inside the wall. The entropy-current matching \(s_+\gamma_+v_+=s_-\gamma_-v_-\) provides an additional condition that closes the system of equations, allowing to estimate the wall velocity --- a key ingredient for the calculation of the gravitational-wave spectrum as it sets the sound shell geometry --- from the pressure \(\mathcal{P}^{\rm LTE}_{\rm bubble}(\xi_w)\) that balances the vacuum driving with the plasma response. 
We extended this approach used for direct transitions towards inverse dynamics, charting the range of transition strength $\alpha_N$ as a function of the enthalpy ratio $\Psi$.
Two generic consequences follow. First, a \emph{nucleation thermodynamic bound} arises from the static limit, \(\mathcal{P}^{\rm LTE}_{\rm bubble}\!\big|_{\xi_w\to 0}=\frac{w_+}{4}(1-\Psi-3\alpha_N)\), so bubbles with \(\alpha_N<(1-\Psi)/3\) (for \(\Psi\equiv a_-/a_+\neq 1\)) are thermodynamically disfavored and cannot expand. 
Second, the terminal speed is set by \(\mathcal{P}^{\rm LTE}_{\rm bubble}(\xi_w)=0\) on the appropriate hydrodynamic branch, with stability controlled by \(d\mathcal{P}^{\rm LTE}_{\rm bubble}/d\xi_w>0\), leading to an upper bound on $\alpha_N$ (depending on $\Psi$) above which the solution enters a runaway regime. 
The resulting phase diagram in the \((\alpha_N,\Psi)\) plane (\cref{Fig:velocity}) exhibits the static-wall boundary, the loci where the LTE pressure vanishes at the direct and inverse Jouguet speeds, and a band of \emph{runaway} solutions within LTE. For inverse phase transitions, transitions with $\alpha_N$ below the slowest inverse hybrid solution with \(\xi_w=c_s^2\), \(\alpha_{N,\rm slowest}\simeq -0.082\), feature a kinematic gap opening around \(v_J^{\rm inv}\) --- i.e.\ a range of wall velocities at fixed $\alpha_N$ in which no solution can be found, as can also be seen in \cref{fig:hydro_solution_space}. 

Independent of the LTE approximation, we computed the sound-wave contribution to the gravitational-wave spectrum from inverse phase transitions using the sound-shell model. 
While the fluid profiles of inverse and direct transitions differ visibly, the corresponding kinetic spectra --- although varying across the respective parameter space --- exhibit similarities between the spectral shapes in direct in inverse hydrodynamics.
The similarities carry over to the present-day gravitational-wave spectrum, rendering a distinction of the underlying hydrodynamics based on an observed gravitational-wave signal difficult.
To further quantify this intuition, we employed two indicators for the discrimination between spectra: the spectral angle --- taking into account only the shape information --- as well as modified version which also accounts for differences in amplitude, confirming that similar signals can be obtained for different parameters across various hydrodynamic branches.
While proper quantitative conclusions require dedicated studies for the corresponding signals in gravitational-wave experiments, our results show that, in order to fully exploit all possible scenarios when attempting to reconstruct phase transition parameters from future observations of a stochastic gravitational-wave background, inverse phase transitions need to be taken into account.

In passing, we have observed that the GW amplitude as predicted within the SSM at low wall velocities but fixed $\alpha_N$ differs significantly from what one would expect by considering the kinetic energy fraction associated to an isolated expanding bubble. This can be traced back to the contribution from the energy contrast, which is not suppressed in the limit of slow walls. It would be interesting to check whether the flattening of the GW peak amplitude for small wall velocities as predicted by the SSM (both for direct and inverse transitions) is actually supported by hydrodynamical simulations. 
This could have important implications for the GW signal from confinement transitions, which may predict slow walls, see e.g.\,\cite{Sanchez-Garitaonandia:2023zqz, Gouttenoire:2023roe}.

To overcome the limitations of the sound-shell model, dedicated 'Higgsless' or field-fluid simulations that more accurately capture the dynamics of inverse phase transitions are required. These simulations would provide a more detailed and nuanced understanding of the gravitational-wave signatures, going beyond the assumptions made in our study. Moreover, further quantitative investigation is needed into the discriminability of inverse phase and direct transitions to properly evaluate the prospects at specific gravitational-wave experiments, particularly in the context of current and future detectors. Lastly, exploring the fundamental model space, including the achievable range of wall velocities and phase transition strengths within physically realistic models, will be essential to assess the true potential for detecting such signatures in the stochastic gravitational-wave background. Such efforts are crucial for fully leveraging gravitational-wave observations to probe the early Universe's phase transition dynamics.

\acknowledgments

It is a pleasure to thank Mark Hindmarsh, Simona Procacci and Alberto Roper Pol for helpful discussions, 
as well as Wenyuan Ai and Benoit Laurent for fruitful comments on the manuscript. 
EM and GB acknowledge support from the Spanish Research Agency (Agencia Estatal de Investigación, MCIN/AEI/10.13039/501100011033) via the IFT Severo Ochoa Center of Excellence grant CEX2020-001007-S. In addition, EM is supported through the grants CNS2023-14453 and PID2022-137127NB-I00 funded by MCIN/AEI/10.13039/501100011033, the European Union NextGenerationEU/PRTR and ESF+, and GB is supported by the grant CNS2023-145069 funded by MICIU/AEI/10.13039/501100011033 and by the European Union NextGenerationEU/PRTR.
M.V. is funded by the European Union (ERC, HoloGW, Grant Agreement No. 101141909). Views and opinions expressed are, however, those of the authors only and do not necessarily reflect those of the European Union or the European Research Council. Neither the European Union nor the granting authority can be held responsible for them. M.V. also acknowledges financial support from Grant CEX2024-001451-M funded by MICIU/AEI/10.13039/501100011033, from Grant No. PID2022-136224NB-C22 from the Spanish Ministry of Science, Innovation and Universities, and from Grant No. 2021-SGR-872 funded by the Catalan Government and by the ``Excellence of Science - EOS'' - be.h project n.30820817, and by the Strategic Research Program High-Energy Physics of the Vrije Universiteit Brussel. SB is supported by the Deutsche Forschungsgemeinschaft under Germany’s Excellence Strategy - EXC 2121 Quantum universe - 390833306.

% >

\appendix

\section{Maximal transition strength}

In this appendix, we briefly delineate the boundaries of the parameter space. For direct deflagrations and hybrid solutions, the transition strength is bounded by \(\alpha_+<\tfrac{1}{3}\). This follows from the requirement that the subsonic branch of the fluid velocity immediately ahead of the wall remains positive, \(v_+^{\text{sub}}(v_-,\alpha_+)>0\), where
\begin{align}
    v_+^\text{sub}(v_-, \alpha_+) = \frac{A(v_-)}{1+\alpha_+} \left[1 - \sqrt{1+ \left(\alpha_++1\right) \left(\alpha_+-\frac{1}{3}\right)/A^2(v_-)}\right] \,, \qquad A \equiv \frac{v_-}{2}+\frac{1}{6 v_-} \,.
\end{align}
For hybrids to evolve consistently toward a shock in front of the wall, one further needs \(\xi_w\,v_+^{\text{sub}}(c_s,\alpha_+)<\tfrac{1}{3}\), which implies
\begin{equation}
\alpha_+>\frac{1-(\sqrt{3}\,\xi_w)^2}{9\,\xi_w^2-1}\,.
\end{equation}
Direct detonations are constrained by the Jouguet condition \(\xi_w>v_J^{\text{direct}}=v_+^{\text{sup}}(c_s,\alpha_+)\), or equivalently
\begin{equation}
\alpha_+<\alpha_J=\frac{(1-\sqrt{3}\,\xi_w)^2}{3(1-\xi_w^2)}\,.
\end{equation}
As a result, there is an overlap between hybrid and detonation branches in the \((\xi_w,\alpha_+)\) plane (hatched region in the right panel of \cref{fig:hydro_solution_space}). This ambiguity disappears if one uses \(\alpha_N\) instead of \(\alpha_+\). For detonations one has \(\alpha_N=\alpha_+\). For hybrids, shock heating ahead of the wall and isentropic processing between shock and wall imply
\begin{align}
    \alpha_N = \alpha_+ \frac{w_+}{w_\text{sh}} \frac{w_\text{sh}}{w_N} > \alpha_+ \frac{w_\text{sh}}{w_N} \,.
\end{align}
Employing the shock condition \(v_{+,\,\text{sh}}\,v_{-,\,\text{sh}}=\tfrac{1}{3}\) with \(v_{+,\,\text{sh}}=\xi_\text{sh}\), the enthalpy jump can be written as
\begin{align}
    \frac{w_\text{sh}}{w_N} = \frac{\gamma_{+,\,\text{sh}}^2 v_{+,\,\text{sh}}}{\gamma_{-,\,\text{sh}}^2 v_{-,\,\text{sh}}} = \frac{9 \xi_\text{sh}^2 -1}{3 (1-\xi_\text{sh}^2)} > \frac{\alpha_J}{\alpha_+^\text{hyb,min}}\,,
\end{align}
where in the last step we used \(\xi_\text{sh}>\xi_w\). Consequently, \(\alpha_N>\alpha_J\) for hybrids, and the pair \((\xi_w,\alpha_N)\) selects a unique solution.

On the inverse side there is no such ambiguity when working with \((\xi_w,\alpha_+)\). The most negative (minimal) strength allowed for inverse deflagrations and inverse hybrids follows from requiring the supersonic branch behind the wall to remain subluminal, \(v_-^{\text{sup}}(v_+,\alpha_+)<1\), with \(v_+=\xi_w>c_s\) for inverse deflagrations and \(v_+=c_s\) for the inverse CJ limit, where
\begin{align}
    v_-^\text{sup/sub}(v_+, \alpha_+) = \frac{1+\alpha_+}{2} v_+ + \frac{1-3 \alpha_+}{6 v_+} \pm \sqrt{\left(\frac{1+\alpha_+}{2} v_+ + \frac{1-3 \alpha_+}{6 v_+}\right)^2 - \frac{1}{3}} \,.
\end{align}
Reaching a shock front in front of the wall further requires \(\xi_w\,v_-^{\text{sub}}(c_s,\alpha_+)>\tfrac{1}{3}\), or equivalently
\[
\xi_w>\frac{1}{3\,v_-^{\text{sub}}(c_s,\alpha_+)}=v_-^{\text{sup}}(c_s,\alpha_+)\equiv v_J^{\text{inv}},
\]
which coincides with the inverse Jouguet velocity and thus marks the boundary of the inverse detonation region. Hence
\[
\alpha_J^\mathrm{inv}=-\frac{\sqrt{3}}{2\,\xi_w}\left(\xi_w-\frac{1}{\sqrt{3}}\right)^2 \, ,
\]
is simultaneously the minimal strength for inverse detonations and the maximal strength for inverse hybrids. The right panel of \cref{fig:hydro_solution_space} displays the resulting domains in the \((\alpha_+,\xi_w)\) plane that admit detonation, deflagration, or hybrid solutions.

Moreover, we can bound the maximal (theoretical) efficiency of kinetic-energy conversion by inspecting the transition’s energy budget. In the direct case, the released vacuum energy and the initial thermal energy are redistributed between bulk fluid motion and the final thermal state, while in the inverse one, the initial energy at our disposal is purely thermal:
\begin{align}
    \textbf{direct}:& \quad \label{eq: energy budget standard}
\underbrace{\frac{\xi_w^3}{3} \epsilon}_{\substack{e_\mathrm{vac}/(4 \pi)\\\text{vacuum energy}\phantom{X}}} +\underbrace{\frac{3}{4} \int w_N \xi^2 \dd \xi}_{\substack{e_\mathrm{ini}/(4 \pi)\\\text{initial thermal energy}\phantom{X}}} = \quad\underbrace{\int \gamma^2 v^2 w \xi^2 \dd \xi}_{\substack{e_K/(4 \pi)\\\text{fluid motion}\phantom{X}}}\quad+ \underbrace{\frac{3}{4} \int w \xi^2 \dd \xi}_{\substack{e_\mathrm{fin}/(4 \pi)\\\text{final thermal energy}\phantom{X}}} \ ,\\
\textbf{inverse}:& \quad \label{eq: energy budget inverse}
\underbrace{\frac{3}{4} \int w_N \xi^2 \dd \xi}_{\substack{e_\mathrm{ini}/(4 \pi)\\\text{initial thermal energy}\phantom{X}}} = \underbrace{\frac{\xi_w^3}{3} \epsilon}_{\substack{e_\mathrm{vac}/(4 \pi)\\\text{vacuum energy}\phantom{X}}} +\quad\underbrace{\int \gamma^2 v^2 w \xi^2 \dd \xi}_{\substack{e_K/(4 \pi)\\\text{fluid motion}\phantom{X}}}\quad+ \underbrace{\frac{3}{4} \int w \xi^2 \dd \xi}_{\substack{e_\mathrm{fin}/(4 \pi)\\\text{final thermal energy}\phantom{X}}} \ .
\end{align}
These balances can be recast as
\begin{align}
     \textbf{direct}:&\quad  \frac{\rho_{\rm kin}}{\rho_{\rm tot}}\equiv \frac{\kappa_{\rm direct}\alpha_N}{1 + \alpha_N}=1-\frac{ \int (w/w_N) \xi^2 \dd \xi }{ 1+\alpha_N} \ \overset{\alpha_N\gg 1}{\underset{w/w_N \to 0}{\longrightarrow}} 1\\
     \textbf{inverse}:&\quad  \frac{\rho_{\rm kin}}{\rho_{\rm tot}}\equiv {\kappa_{\rm inverse}}=1-|\alpha_N|-\frac{\frac{3}{4} \int w \xi^2 \dd \xi}{\frac{3}{4} \int w_N \xi^2 \dd \xi} ~~ \overset{\alpha_N\to -1/3}{\underset{w/w_N \to 0}{\longrightarrow}} ~~2/3 \ ,
\end{align}
where the limiting behaviors highlight the maximal conversion consistent with energy conservation. 
Here, $\kappa_\mathrm{direct} = e_K/e_\mathrm{vac}$~\cite{Espinosa:2010hh} and $\kappa_\mathrm{inverse} = e_K/e_\mathrm{ini}$~\cite{Barni:2024lkj}. In the direct case, for very strong transitions ($\alpha_N\gg1$) the kinetic-energy fraction can approach unity. In the inverse case, even in the idealized limit where the initial thermal reservoir is converted only into vacuum energy and bulk motion ($\alpha_N\to-1/3$~\cite{Barni:2024lkj}), the maximal fraction is bounded by $2/3$. This explains why complete efficiency is unattainable for inverse transitions.

\section{Connection between single-bubble and multi-bubble: RMS velocity and conversion factors }
\label{app:rms}

In this appendix we bridge the single-bubble hydrodynamic picture with the ensemble description of the multi-bubble sound field.
While both are governed by the same underlying fluid dynamics, they differ in the type of averaging involved: the single-bubble treatment describes the deterministic self-similar flow around one expanding bubble, whereas the multi-bubble regime corresponds to the stochastic superposition of many such profiles nucleated at random times and positions including the dynamics after the bubbles have merged and the transition is complete.
The distinction between these two levels of description emerges when computing quadratic observables such as the kinetic-energy density or the enthalpy-weighted RMS velocity: in the single-bubble case these follow directly from the Fourier kernel $f'(z)$, whereas in the multi-bubble field they involve its acoustic continuation $A(z)$ and statistical factors that encode the distribution of bubbles.
By tracing this connection step by step, following the derivation in Ref.~\cite{Hindmarsh:2019phv}, we reconcile the definitions of kinetic-energy conversion and RMS velocity used in the Sound Shell Model with those derived from single-bubble hydrodynamics.

\subsection{Linearised perturbations}
After bubble collisions, the subsequent fluid motion is well described by small, longitudinal perturbations that
propagate as linear sound waves. In this regime we track the coupled evolution of the
energy-density fluctuation and the velocity field. For non-relativistic velocities we define
$
\lambda(\mathbf x,t)\equiv \frac{\rho(\mathbf x,t)-\bar \rho}{\bar w},
$
and the linearised equations in Fourier space read as in \cref{eq:ssm:lin-eqs}. Writing the general longitudinal solution as a
superposition of plane waves,
\begin{equation}
u^i(\mathbf x,t)=\int\!\frac{\dd^3q}{(2\pi)^3}\,
\hat q^i\Big(u({\mathbf q})\,e^{-i\omega t+i\mathbf q\cdot\mathbf x}
+ u^{*}({\mathbf q})\,e^{+i\omega t-i\mathbf q\cdot\mathbf x}\Big),
\end{equation}
the plane-wave amplitude at the initial time $t_i$ is obtained by Fourier transforming the
fields and using \cref{eq:ssm:lin-eqs}. One finds
\begin{equation}
u({\mathbf q})
=\frac{1}{2}\Big(\hat q_i\, u^i({\mathbf q}(t_i))-c_s\,\lambda({\mathbf q}(t_i))\Big)
\,e^{\,i\omega t_i},
\end{equation}
which expresses the freely propagating sound mode in terms of the longitudinal velocity
and the enthalpy perturbation evaluated at $t_i$.

Before collisions, each bubble generates a spherically-symmetric self-similar
fluid profile of the form
\begin{equation}
v^{(n)}_i(\mathbf x,t)
= \hat r_i^{(n)}\,v_{\rm ip}(\xi), 
\qquad 
\xi \equiv \frac{r}{T^{(n)}},
\end{equation}
where $\mathbf{r}^{(n)} = \mathbf{x}-\mathbf{x}^{(n)}$ is the radial vector between the $n$-th bubble center and the position $\mathbf{x}$, $T^{(n)}$ is the time since
nucleation of the $n$-th bubble, and $v_{\rm ip}(\xi)$ is the invariant velocity
profile (the subscript ``ip'' stands for invariant profile).
Because the system is self-similar, the spatial dependence enters only through
the ratio $\xi = r^{(n)}/T^{(n)}$.

The Fourier transform of this velocity field is defined as\footnote{We are using two different symbols for the velocity: $v$ we refer to the single-bubble velocity, while $u$ is meant for the velocity in the multi-bubble case.}
\begin{equation}
 v^{(n)}_i(\mathbf q,t)
\equiv \int\! \dd^3x \, e^{-i\mathbf q\cdot \mathbf x}\, v^{(n)}_i(\mathbf x,t)
= \int\! \dd^3x \, e^{-i\mathbf q\cdot \mathbf x}\, \hat r_i^{(n)}\,v_{\rm ip}(\xi) .
\label{eq:fourier_def}
\end{equation}
Because the velocity is purely longitudinal, its transform must also be proportional to the
unit vector $\hat q_i \equiv q_i/q$ in momentum space. This implies that
\begin{equation}
 v^{(n)}_i(\mathbf q,t)
= \hat q_i\,  v_L^{(n)}(q,t),
\end{equation}
where $ v_L^{(n)}(q,t)$ is the longitudinal scalar amplitude.
To obtain this amplitude explicitly, we can project \cref{eq:fourier_def}
along $\hat q_i$ and use spherical coordinates with the $\hat q$-axis as polar axis.
To further make the self-similar scaling explicit, change variables to
$\xi = r^{(n)}/T^{(n)}$ and define the dimensionless wave number $z = q T^{(n)}$, therefore
\begin{align}
 v_L^{(n)}(q,t)
&= 4\pi \big(T^{(n)}\big)^3  e^{-i\mathbf q\cdot \mathbf x^{(n)}}
\int\! \dd\xi\, \xi^2\, v_{\rm ip}(\xi)
\frac{\sin(z\xi)-z\xi\cos(z\xi)}{(z\xi)^2}.
\label{eq:vL_integral}
\end{align}
Now, with the auxiliary function defined in \cref{eq:ssm:f-l-kernels}, the integral \cref{eq:vL_integral} can be expressed compactly and restoring the vector structure we obtain
\begin{equation}
 v^{(n)}_i(\mathbf q,t)
= e^{-i\mathbf q\cdot \mathbf{x}^{(n)}}\,
\big(T^{(n)}\big)^3\,
\hat q_i\, f'(z) .
\label{eq:single bubble velocity Fourier}
\end{equation}
\subsection{General definition of the RMS velocity}

We define the enthalpy-weighted RMS four-velocity as
\begin{equation}
\bar{U}_f^{\,2}
\;\equiv\;
\frac{\langle w\,\gamma^2 v^2\rangle}{\langle w\rangle}
\label{eq:defU}
\end{equation}
Introducing the enthalpy-weighted momentum field
\( \mathbf b(\mathbf x)\equiv\sqrt{w(\mathbf x)}\,\gamma(\mathbf x)\,\mathbf v(\mathbf x) \),
we can express the average $\langle w\,\gamma^2 v^2\rangle$ in the denominator of \cref{eq:defU} in terms of $|\mathbf{b}|^2$, applying Parseval's identity alternatively in position or momentum space, as
\begin{equation}
V\, \langle w\,\gamma^2 v^2\rangle = \int \!\dd^3x\,|\mathbf b(\mathbf x)|^2
= \int\!\frac{\dd^3q}{(2\pi)^3}\,|{\mathbf b}(\mathbf q)|^2,
\label{eq:parseval}
\end{equation}
where $V$ is the averaging volume.

\paragraph{Single-bubble average.}
For one isolated bubble, the enthalpy-weighted RMS fluid velocity is usually obtained normalizing the total kinetic energy $e_K \equiv 4 \pi \int_0^{\xi_{\max}}\dd\xi\,\xi^2 w\,\gamma^2 v^2$, where $\xi_{\max}=\max(\xi_w,\xi_\mathrm{sh},c_s)$ is the largest self-similar coordinate with non-vanishing fluid velocity, to the enthalpy within a bubble volume $V_\mathrm{bubble} = 4 \pi \xi_w^3/3$ in the symmetric phase,
\begin{equation}
    \label{eq:Ubar_single_position}
    \bar{U}_{f,\text{single}}^{\,2} = \frac{e_K}{V_\mathrm{bubble} \bar{w}}
=\frac{\int\dd^3x\, |\mathbf{b}^2(\mathbf{x})|}{V_\mathrm{bubble}\, \bar{w}}
= \frac{1}{\bar{w}}\,\frac{3}{\xi_w^3}\int_0^{\xi_{\max}} d\xi\,\xi^2\,w\,\gamma^2 v^2
\,.
\end{equation}
The velocity field is radial and self-similar,
\( v_i(\mathbf x,t)=\hat r_i\,v_{\rm ip}(\xi) \) with \( \xi=r/T \).
The corresponding enthalpy-weighted momentum field can be written as
\begin{equation}
b_i(\mathbf x,t)=\sqrt{w(\mathbf x,t)}\,\gamma(\mathbf x,t)\,v_i(\mathbf x,t)
=\hat r_i\,\sqrt{w(\xi)}\,\gamma(\xi)\,v_{\rm ip}(\xi),
\end{equation}
which depends on space only through the self-similar coordinate $\xi$.
Following the same steps leading to \cref{eq:vL_integral}, its Fourier transform is given by
\begin{align}
b_i(\mathbf q,t)
&=\sqrt{\bar w}\,T^3\,\hat q_i\,\phi'(z)\,,
&
\phi'(z) &\equiv {4\pi\int d\xi\,\xi^2\,
\sqrt{\dfrac{w(\xi)}{\bar w}}\,\gamma(\xi)\,v_{\rm ip}(\xi)\,
\frac{\sin(z\xi)-z\xi\cos(z\xi)}{(z\xi)^2}}
\,,
\end{align}
where $z=q T$.
The dimensionless kernel $\phi'(z)$ generalizes the velocity kernel $f'(z)$ by including the enthalpy weight.
Applying Parseval’s identity, \cref{eq:parseval}, we can rewrite \cref{eq:Ubar_single_position} in terms of the kernel as
\begin{equation}
\bar{U}_{f,\;{\rm single}}^{\,2}
=\frac{3}{4\pi \xi_w^3}\int_0^\infty\frac{dz}{(2\pi)^2}\,z^2\,|\phi'(z)|^2\,.
\label{eq:U_single}
\end{equation}

\paragraph{Multi-bubble average.}
In the multi-bubble regime the plasma velocity is a random superposition of the flows
generated by individual bubbles nucleated at different spacetime points,
\begin{equation}
u_i(\mathbf x,t)=\sum_n u^{(n)}_i(\mathbf x,t),
\qquad
b_i(\mathbf x,t)=\sqrt{w}\,\gamma\,u_i=\sum_n b^{(n)}_i(\mathbf x,t),
\end{equation}
where $b_i$ is the enthalpy-weighted velocity field.
In Fourier space,
\begin{equation}
 b_i(\mathbf q,t)=\sum_n  b^{(n)}_i(\mathbf q,t).
\end{equation}

After bubble overlap, the sound waves evolves as free perturbation and the of the $b$ field can be derived as in \cref{eq:ssm:Apm}. Now since we will compute the multi-bubble average of $b_i(\mathbf x,t)=\sqrt{w}\,\gamma\,u_i$ we need to introduce the amplitude $\mathcal{B}_\pm(z)$ of such field in Fourier space, so that the single-bubble Fourier modes read
\begin{equation}
b^{(n)}_i(\mathbf q,t)
=\bar w\,(T^{(n)})^3\,\hat q_i
\Big[\mathcal B_+(z_n)\,e^{-i\omega t}
+\mathcal B_-(z_n)\,e^{+i\omega t}\Big]
e^{-i\mathbf q\cdot\mathbf x^{(n)}}
\qquad\text{with}\qquad
\mathcal{B}_\pm (z)\equiv \frac{1}{2}\big(\phi'(z)\pm i\,c_s\,\eta(z)\big),
\label{eq:mathcalA_def}
\end{equation}
where $\eta(z)$ is the scalar kernel built from the weighted
enthalpy perturbation,
\begin{equation}
\eta(z)=4\pi\!\int_0^\infty\! d\xi\,\xi^2\,
\sqrt{\dfrac{w(\xi)}{\bar w}}\;\lambda_{\rm ip}(\xi)\,
\frac{\sin(z\xi)}{z\xi}.
\end{equation}
In the limit of $\xi_w \ll 1$, we notice that $\lambda_{\rm ip}(\xi)$ is basically a constant $\mathcal{O}(\alpha_N)$ up to $\xi = \xi_w$, namely inside the bubble, and approximately zero outside. For this reason, the contribution from the energy contrast will be the dominant one at small wall velocity, as we shall see. 

The total Fourier amplitude is the sum over all bubbles,
$ b_i=\sum_n b_i^{(n)}$, so that
\begin{equation}
\Big\langle b_i(\mathbf q,t)\, b_i^*(\mathbf q,t)\Big\rangle
=\sum_{n,m}
\Big\langle b_i^{(n)}(\mathbf q,t)\,
 b_i^{(m)*}(\mathbf q,t)\Big\rangle.
\end{equation}
Averaging over random centers $\mathbf x^{(n)}$ removes the interference
terms ($n\!\ne\! m$) for $\mathbf q\!\ne\!0$ through phase cancellation
$e^{-i\mathbf q\cdot(\mathbf x^{(n)}-\mathbf x^{(m)})}$.
A further average over initial times $t_i^{(n)}$ makes the two acoustic branches
orthogonal, yielding
\begin{equation}
\big\langle|\mathcal B_+|^2+|\mathcal B_-|^2\big\rangle
=2\,|\mathcal B(z)|^2.
\end{equation}
\paragraph{Average over bubble ages.}
At this stage, we have averaged over the random spatial positions of bubbles, which removes cross-terms between different bubbles. The next step is to average over their \emph{ages}, since bubbles nucleated at different times contribute with different radii $T^{(n)}$ and hence different self-similar variables $z_n=qT^{(n)}$.  

Instead of summing explicitly over all bubbles, it is convenient to describe the population statistically. As in \cref{eq:ssm:nucleation-weights}, let $\nu(T)$ be the normalized age distribution and $\mathcal N$ the bubble number density. The discrete sum over bubbles is then replaced by an ensemble integral,
\begin{equation}
\sum_n \langle \cdots \rangle
\;\longrightarrow\;
\mathcal N\,V\!\int dT\,\nu(T)\,(\cdots),
\end{equation}
which expresses that, in a homogeneous stochastic ensemble, the average over many bubbles is equivalent to integrating over their age distribution.

Since each single-bubble amplitude scales as $T^3$ and depends on $z=qT$, changing variables $T\!\to\!z$ introduces a finite statistical factor,
\begin{equation}
\nu_3\;\equiv\;\int dT\,\nu(T)\,T^3,
\end{equation}
corresponding to the third moment of the age distribution. This factor captures the average geometric weight of bubbles at the observation time and determines the overall normalization of the ensemble average, while the spectral shape remains encoded in $|\mathcal B(z)|^2$. For standard nucleation scenarios (exponential or simultaneous) one finds $\nu_3=6$.

Using Parseval’s identity \cref{eq:parseval} and integrating over all
modes ($d^3q=4\pi q^2dq$), the averaged spectrum becomes
\begin{equation}
\int\!\frac{d^3q}{(2\pi)^3}\,
\Big\langle|{\mathbf b}(\mathbf q,t)|^2\Big\rangle
=\frac{\bar w\,\nu_3}{(2\pi)^2}\!\int_0^\infty\!dz\,z^2\,2|\mathcal B(z)|^2.
\end{equation}
Equating this with the real-space definition of the RMS,
\begin{equation}
\frac{1}{V}\int d^3x\,|\mathbf b|^2
=\frac{1}{V}\int d^3x\,w\,\gamma^2v^2
=\frac{3}{4\pi \xi_w^3}\int_0^{\xi_{\max}}d\xi\,\xi^2w\,\gamma^2v^2,
\end{equation}
and dividing by $\bar w$ finally yields the enthalpy-weighted
RMS four-velocity of the multi-bubble sound field:
\begin{equation}
\bar{U}_{f,{\rm multi}}^{\,2}
=\frac{1}{\bar w}\,\frac{1}{V}\int d^3x\,w\,\gamma^2v^2
=\frac{\nu_3}{4\pi \xi_w^3}
\int_0^\infty\!\frac{dz}{(2\pi)^2}\,z^2\,2\,|\mathcal B(z)|^2\;.
\label{eq:U_multi}
\end{equation}
This expression is the exact enthalpy-weighted counterpart of the
standard SSM formula, obtained by replacing
$f'(z)\!\to\!\phi'(z)$ and $\ell(z)\!\to\!\eta(z)$ in the acoustic amplitude.

\subsection*{Relation between \texorpdfstring{$\Omega_K$}{the kinetic energy fraction} and the multi-bubble RMS velocity}

To connect the multi-bubble result with single-bubble hydrodynamics, it is useful to decompose the total RMS velocity into two distinct pieces:
\begin{align}
    \bar U_{f, ~ {\rm multi}}^2 &= \frac{3 }{ 4\pi \xi_w^3}\int_0^\infty \frac{d z}{(2\pi)^2}\,z^2\Big[ |\phi'(z)|^2+c_s^2|\eta(z)|^2\Big] \\
    &=\bar U_{f, ~ {\rm single}}^2+\frac{3}{4\pi \xi_w^3}\int_0^\infty \frac{d z}{(2\pi)^2}\,z^2\,c_s^2|\eta(z)|^2 \, .
\end{align}
This split makes the physical interpretation transparent: the first term corresponds to the single-bubble contribution, while the second arises from the collective, enthalpy-weighted overlap of bubbles after collision. The second term is strictly positive, reflecting the fact that the linear acoustic perturbations generated during collisions always add extra kinetic energy. One can show that the second piece goes to a constant $\mathcal{O}(\alpha_N^2)$ for small wall velocity, $\xi_w \rightarrow 0$, so that in this limit this dominates over the first term, which instead vanishes for $\xi_w \rightarrow 0$.

It follows that the kinetic-energy fraction can be written as
\begin{equation}
    \Omega_K=\Gamma\,\bar U_{f, ~ {\rm multi}}^2 =\Omega_K^{\rm single}+\Omega_K^{\rm overlap} >\Omega_K^{\rm single}\,,
\end{equation}
where the inequality simply expresses that multi-bubble overlap enhances the total kinetic fraction beyond the single-bubble limit. Physically, the first term represents the energy stored in the fluid surrounding an isolated expanding wall, while the second term quantifies the contribution from overlapping sound shells under the assumption of linear post-collision evolution.

To connect $\Omega_K$ with the efficiency factors used in the direct and inverse hydrodynamic regimes, and following Ref.~\cite{Barni:2024lkj}, we recall the definitions
\begin{equation}
    \kappa_{\rm direct}= \frac{e_K}{\frac{4\pi}{3}\xi_w^3 \epsilon}, \qquad 
    \kappa_{\rm inverse}= \frac{ e_K }{ 4 \pi \int_0^{\xi_{\rm max}}d\xi \,\xi^2 \,\frac{3}{4}\,w_N}\,,
\end{equation}
where $e_K$ is the bulk kinetic energy, $\epsilon$ is the $T=0$ vacuum energy, and $w_N$ the enthalpy of the old phase. Using these, the kinetic fraction in the sound-shell model reads
\begin{align}
    \Omega_K^{\rm direct}&= \alpha_N\,\kappa_{\rm direct}
    +\frac{3\Gamma}{4\pi \xi_w^3}\int_0^\infty \frac{dz}{(2\pi)^2}\,z^2\,c_s^2|\eta(z)|^2\,,
    \label{eq:OmegaK_kappa_direct}\\[4pt]
    \Omega_K^{\rm inverse}&= \frac{3}{4 \pi \xi_w^3}\,\kappa_{\rm inverse}
    +\frac{3\Gamma}{4\pi \xi_w^3}\int_0^\infty \frac{dz}{(2\pi)^2}\,z^2\,c_s^2|\eta(z)|^2\,.
    \label{eq:OmegaK_kappa_inverse}
\end{align}
In the direct case, $\Omega_K^{\rm single}$ corresponds to the usual efficiency times the released-energy fraction, while in the inverse case it scales with the geometric factor $(4\pi/3)\xi_w^3$.

\begin{figure}[t]
  \centering
  \includegraphics[width=0.49\linewidth]{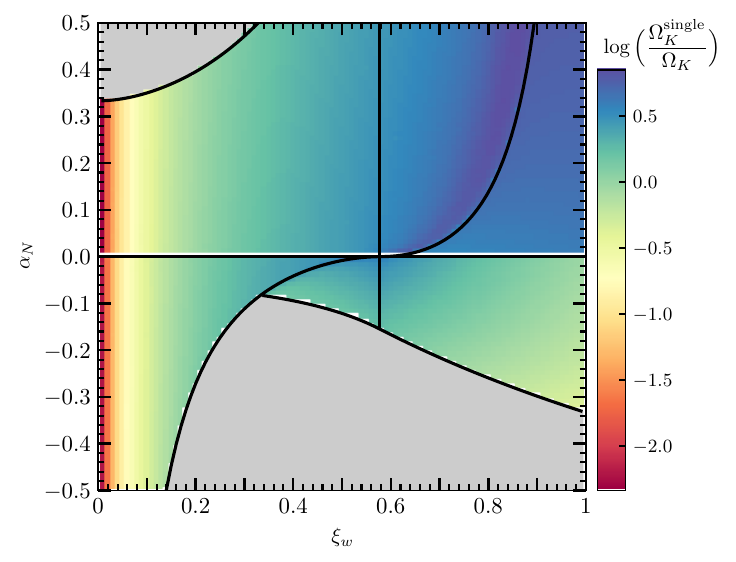}%
  \caption{Normalized contrast maps of single- vs.\ multi-bubble efficiencies over $(\xi_w,\alpha_N)$, showing $\log\!\big(\Omega_K^{\rm single}/\Omega_K\big)$. Red regions mark strong discrepancies at low wall velocities, where single-bubble conversion is inefficient while the multi-bubble kinetic fraction remains sizeable due to the accumulation of overlapping acoustic shells.}
  \label{fig:OmegaK_vs_kappa}
\end{figure}

The relations above make it clear that $\Omega_K$ receives two distinct contributions: 
a first term associated with the single-bubble efficiency (the part directly linked to $\Omega_K^{\rm single}$), 
and a second term emerging from the enthalpy-weighted energy contrast of the overlapping shells. 
When the wall velocity decreases, the single-bubble contribution vanishes rapidly—this reflects the steep drop of the kinetic energy generated by an isolated expanding wall. 
In contrast, the enthalpy-weighted term, which originates from the collective overlap of bubbles and from the assumption of linear acoustic evolution after collision, 
remains finite at low $\xi_w$. 
Intuitively, this can be understood by noting that, for a fixed energy contrast $\alpha_N$, 
the global enthalpy difference between phases persists even if the walls move slowly: 
the plasma still carries pressure and enthalpy perturbations that continue to drive acoustic motion. 
Hence, the multi-bubble component encoded in \cref{eq:U_multi} dominates at small wall velocities, 
capturing the sustained coherent motion that the single-bubble efficiency alone cannot describe.

This explains the behavior observed in \cref{fig:OmegaK_vs_kappa},
which shows the logarithmic contrast between $\Omega_K^\mathrm{single}$ and $\Omega_K$ across the $(\xi_w,\alpha_N)$ plane. 
Red regions indicate where $\Omega_K^\mathrm{single}\!\ll\!\Omega_K$, i.e.\ where local conversion is inefficient but collective perturbations remain significant. 

\bibliography{biblio}

\end{document}